\documentclass[aps, prd, superscriptaddress , amsmath, amssymb, twocolumn, nofootinbib]{revtex4-1}

\usepackage[T1]{fontenc}
\usepackage{amsmath,amsfonts,bbm, graphicx, color}
\usepackage{perpage}
\MakePerPage{footnote}
\usepackage{amssymb}
\usepackage{appendix}
\usepackage{relsize}
\usepackage[colorlinks]{hyperref}
\usepackage[figure,table]{hypcap}
\usepackage[textsize=tiny]{todonotes}
\usepackage{mathtools}

\hypersetup{
	bookmarksnumbered,
	pdfstartview={FitH},
	citecolor={green},
	linkcolor={darkblue},
	urlcolor={darkblue},
	pdfpagemode={UseOutlines}}
\definecolor{darkgreen}{RGB}{20,100,20}
\definecolor{darkblue}{RGB}{0,0,130}
\definecolor{darkred}{rgb}{.8,0,0}

\def\>{\rangle}
\def\<{\langle}

\def\N{ {\cal N} }

\def\A{ {\cal A} }

\def\>{\rangle}
\def\<{\langle}
\newcommand{\ket}[1]{|#1 \rangle}
\newcommand{\bra}[1]{\langle #1|}

\newcommand{\bi[1]}{\hat{b}_{\text{I}#1}}

\newcommand{\bii[1]}{\hat{b}_{\text{II}#1}}
\newcommand{\wi[1]}{w_{\text{I}#1}}
\newcommand{\wii[1]}{w_{\text{II}#1}}
\newcommand{\psii}{\psi_\text{I}}
\newcommand{\psiii}{\psi_\text{II}}
\newcommand{\phii}{\phi_\text{I}}
\newcommand{\phiii}{\phi_\text{II}}
\newcommand{\ali}{\alpha_\text{I}}
\newcommand{\alii}{\alpha_\text{II}}
\newcommand{\bei}{\beta_\text{I}}
\newcommand{\beii}{\beta_\text{II}}

\newcommand{\oI}{(\psi_\text{I}, w_{\text{I}\Omega{\bf k_\perp}})}
\newcommand{\oIl}{(\psi_\text{I}, w_{\text{I}\Xi{\bf l_\perp}})}
\newcommand{\oII}{(\psi_\text{II}, w_{\text{II}\Xi{\bf l_\perp}})}
\newcommand{\oIIk}{(\psi_\text{II}, w_{\text{II}\Omega{\bf k_\perp}})}
\newcommand{\oIIlk}{(\psi_\text{II}, w_{\text{II}\Xi{\bf k_\perp}})}
\newcommand{\oIImk}{(\psi_\text{II}, w_{\text{II}\Omega{\bf -k_\perp}})}
\newcommand{\oIIlmk}{(\psi_\text{II}, w_{\text{II}\Xi{\bf -k_\perp}})}

\def\wiOk{w_{\text{I}\Omega{\bf k_\perp}}}
\def\wiiOk{w_{\text{II}\Omega{\bf k_\perp}}}

\def\biOk{\hat{b}_{\text{I}\Omega{\bf k_\perp}}}

\def\biiOk{\hat{b}_{\text{II}\Omega{\bf k_\perp}}}

\newcommand{\oIokL}{(\psi_\Lambda, w_{\Lambda\Omega{\bf k_\perp}})}

\newcommand{\pa}{{\mkern3mu\vphantom{\perp}\vrule depth 0pt\mkern2mu\vrule depth 0pt\mkern3mu}}
\newcommand{\pe}{\perp}

\def\d{^{\dag}}
\def\s{^{\star}}
\renewcommand{\Re}{\text{Re}} 
\renewcommand{\Im}{\text{Im}}

\def\I{ \mathbbm{1} }

\def\>{\rangle}
\def\<{\langle}
\def \be{\begin{equation}}
\def \ee{\end{equation}}
\def \beq{\begin{equation}}
\def \eeq{\end{equation}}
\def \bea{\begin{eqnarray}}
\def \eea{\end{eqnarray}}

\renewcommand{\[}{\begin{equation}}
\renewcommand{\]}{\end{equation}}

\renewcommand{\#}{,\ \;}

\definecolor{darkred}{rgb}{.8,0,0}
\definecolor{magenta}{RGB}{255,0,255}
\definecolor{green}{rgb}{.2,.6,.2}

\usepackage{soul}
\newcommand{\nn}{\nonumber}

\begin{document}

\title{Two-mode Gaussian quantum states measured by collinearly and noncollinearly accelerating observers}

\author{Piotr T. Grochowski}
\email{piotr@cft.edu.pl}
\affiliation{Institute of Theoretical Physics, University of Warsaw, Pasteura 5, 02-093 Warsaw, Poland}
\affiliation{Center for Theoretical Physics of the Polish Academy of Sciences, Aleja Lotnik\'ow 32/46, 02-668 Warsaw, Poland}
\author{Krzysztof Lorek}
\email{krzysztof.lorek@fuw.edu.pl}
\affiliation{Institute of Theoretical Physics, University of Warsaw, Pasteura 5, 02-093 Warsaw, Poland}
\author{Andrzej Dragan}
\email{dragan@fuw.edu.pl}
\affiliation{Institute of Theoretical Physics, University of Warsaw, Pasteura 5, 02-093 Warsaw, Poland}
\affiliation{Centre for Quantum Technologies, National University of Singapore, 3 Science Drive 2, 117543 Singapore, Singapore}
\date{\today}

\begin{abstract}
We generalize $1+1$-dimensional formalism derived by Ahmadi et. al. [Phys. Rev. D \textbf{93}, 124031] to investigate an effect of relativistic acceleration on localized two-mode Gaussian quantum states in $3+1$-dimensional spacetime.
The following framework is then used to analyze entanglement of the Minkowski vacuum as witnessed by two accelerating observers that move either collinearly or noncollinearly. 
\end{abstract}


\maketitle

\section{Introduction}
Since 1970s it is known that the concept of a particle in quantum field theory is observer-dependent~\cite{Fulling1973a,Hawking1974,Hawking1975,Unruh1977,Davies1975a}.
In 1976 Unruh showed that a uniformly accelerated particle detector in the vacuum perceives a thermal bath of particles~\cite{Unruh1976a} -- the effect that is now known by his name~\cite{Takagi1986,Crispino2008}.
The Unruh effect is however not just an odd curiosity, arising from some peculiar mathematical considerations.
It pertains to many basic concepts in fundamental physics -- among the others, entanglement~\cite{,Alsing2012}, black hole thermodynamics~\cite{Sciama1981}, and Einstein equations as equations of state for spacetimes in thermal equilibrium~\cite{Jacobson1995}.
More detailed descriptions of these and more aspects of the Unruh effect have been gathered in many reviews, including~\cite{,DeWitt1975,Davies1978,Sciama1981,Birrell1982,Takagi1986,Fulling1987,Ginzburg1987,Wald1994,Crispino2008}.

The connection between quantum mechanics, relativity and theory of information have been investigated since the advent of these early observations, but it is still a rapidly developing research area~\cite{Peres2004,Alsing2012}.
First relativistic considerations of quantum information were done by Czachor in 1997, who provided a relativistic background to the well-known Einstein-Podolski-Rosen-Bohm experiment~\cite{,Alsing2002,TERASHIMA2003,Peres2004,Czachor1997,Ahn2003,Massar2006,Ball2006a}.
It was not until 2002 when the first works on quantum entanglement in the presence of spacetime appeared~\cite{Peres2004,Peres2002,Alsing2002,Gingrich2002,Gingrich2003,Pachos2002}.

Early results involving accelerating observers suggested that entanglement is indeed observer-dependent~\cite{,Alsing2003b,Alsing2004,Fuentes-Schuller2005a,Alsing2006a,Bruschi2010,Bruschi2012b}.
Pioneering work of Alsing and Milburn~\cite{Alsing2003b} considered a teleportation protocol performed by inertial Alice and her uniformly accelerating partner Rob.
The fidelity was compromised in comparison to the inertial scenario, strongly suggesting degradation of entanglement that goes beyond Lorentz mixing of degrees of freedom. 
This result was later confirmed by a study of the entanglement of two field modes in a similar setup~\cite{,Fuentes-Schuller2005a}.
Moreover, works by Reznik \textit{et al.} showed that spatial degrees of freedom of global modes are also entangled, including even the vacuum state~\cite{,Reznik2003,Reznik2005,Lin2010,Olson2011}.

Unfortunately, most of these early studies on entanglement in non-inertial frames considered only global modes, that are not well-suited for quantum-information protocols.
For a more realistic setup in which quantum states can be measured, transferred and exploited, some kind of localization in space and time is necessary.
Different approaches have been employed to tackle this problem -- moving cavities~\cite{,Bruschi2012a,Friis2012,Friis2012a,Bruschi2013a,Friis2012b}, point-like detectors~\cite{Lin2008,Lin2008a,Lin2009,Lin2015,Doukas2010} and localized wave-packets~\cite{Downes2013,Dragan2012,Dragan2013b,Doukas2013a,Richter2017,Ahmadi2016a,Grochowski2017,Debski2018}.

In this work, we focus on the latter approach as we generalize the framework established in Ref.~\cite{Ahmadi2016a} that introduces a way to compute the effect of acceleration on two-mode Gaussian states of $1+1$-dimensional localized wavepackets.
We investigate a similar Gaussian channel, however in the $3+1$-dimensional spacetime.
Such a extension allows us to study more complicated geometries, involving analysis of the role of perpendicular spatial degrees of freedom and relative rotation of the trajectories of involved observers.
Beside the explicit calculation of the parameters of the channel, we also discuss the amount of entanglement that a pair of accelerated observers can witness while moving in the vacuum state.
We have further confirmed results obtained previously in the literature, and have gone beyond by checking that the correlations between observers can increase if they move non-collinearly.

The paper is organized as follows.
In Section~\ref{theframework}, we briefly reintroduce a quantum Gaussian channel from~\cite{Ahmadi2016a} and generalize it in order to describe $3+1$-dimensional states.
This part introduces all the necessary notions that are relevant for later considerations.
In Section~\ref{compchannel}, we provide explicit expressions for the parameters of the channel in varying geometries, leaving meticulous computations out in Appendix~\ref{computingnoise}.
Section~\ref{modechoice} discusses the choice of the wave packets that observers have access to.
In Section~\ref{vacent} the amount of entanglement witnessed by two collinearly accelerating observers is numerically evaluated.
The penultimate Section~\ref{r:2d} treats a noncollinear scenario -- contrary to previous Sections, in there we consider $2+1$-dimensional spacetime to simplify difficult computations.
We derive a quantum channel that describes this case and provide numerical calculations for a small relative angle between the observers.
The quantity we investigate is again the vacuum entanglement.
Detailed calculations from this Section are provided in Appendix~\ref{computingnoise2}.
Finally, we finish our work with conclusions and the outlook for the future research in Section~\ref{concl}. 

\section{The framework\label{theframework}}
The aim of the following Section is to briefly recapitulate a quantum channel that accounts for acceleration effects on two-mode Gaussian states.
Such a framework was presented in Ref.~\cite{Ahmadi2016a} and for a detailed introduction, see Section II there.
Our main purpose is to generalize this approach into $3+1$-dimensional spacetime and as such, we will focus on underlining roles of similarities and differences that arise while changing the dimensionality.
Throughout this work we use natural units with $c=\hbar=1$.
\subsection{Outline}
We investigate a real bosonic field $\hat{\Phi}$ with a mass $m$, in a $3+1$-dimensional Minkowski spacetime.
In contrast to $1+1$-dimensional case, there is no problem with the infrared divergence -- the limiting case $m \rightarrow 0$ can be calculated just by putting $m=0$.
The evolution of the field is governed by the Klein-Gordon equation, $(\Box+m^2)\hat{\Phi}=0$ that implies a canonical scalar product:
\begin{align}\label{SP}
(\phi_1,\phi_2) &= i\int_{\Sigma}\mbox{d}^3{\bf x} \left(\phi_1^\star \partial_t\phi_2-\phi_2\partial_t\phi_1^\star \right),
\end{align}
where $\Sigma$ is a spacelike Cauchy surface.

We will investigate a transformation of a Gaussian state of two wavepackets stationary in the inertial frame, into a Gaussian state of two uniformly accelerated wavepackets.
This will involve two decompositions of the field $\hat{\Phi}$ into two sets of orthonormal modes with respect to the Klein-Gordon inner product~\eqref{SP}.
The inertial modes will be denoted as $\phi_n$, with associated annihilation operators $\hat{f}_n$ which consist only of positive frequencies with respect to the Minkowski timelike Killing vector field.
The accelerated modes will be denoted as $\psi_n$, with associated annihilation operators $\hat{d}_n$ which consist only of positive frequencies with respect to the Rindler timelike Killing vector field.
These two decompositions of the quantum field can be written as 
\begin{equation}\label{phipsiexpa}
\hat{\Phi}=\sum_n\phi_n\hat{f}_n+\text{H.c.}=\sum_n\psi_n\hat{d}_n+\text{H.c.}
\end{equation}

The demand for the lack of the negative frequency contribution is due to the construction of the relativistic Glauber detector~\cite{Dragan2012}.
If a wavepacket possessed such a contribution, the detector would experience so called \textit{dark counts} -- it would click even in the vacuum state.

In our setup, we choose two modes, $\phii$ and $\phiii$, out of the orthonormal set of the inertial modes.
They will be prepared in a certain two-mode Gaussian state, while the remaining modes in the set will be in the vacuum state.
Out of the set of accelerated modes, $\psii$ and $\psiii$ will be associated with the accelerating observers, while the remaining ones, although not empty, will be traced out.
Ignoring these modes will be the cause for the Gaussian noise in the state transformation from the inertial frame to the accelerating frame. 
The goal of this work is to derive an expression for a quantum channel, transforming the state of $\phii$ and $\phiii$ into the state of $\psii$ and $\psiii$.

In practice one usually deals with localized states, and most certainly an observer or a rigid detecting device is localized.
Therefore one can associate a single proper acceleration with it, which we identify as a proper acceleration of the center of the wave packet.
For this reason we choose the $\phi_n$'s and $\psi_n$'s to be localized in all three spatial dimensions.
However, since they consist of only positive frequencies, their support in the position space has to be noncompact.
The localization is therefore not strict, i.e. the modes are allowed to have infinite tails.
Our approach is based on the choice of a specific spatial envelope of the modes, that by a sufficiently fast decay guarantees that the negative-frequency contribution is negligible.
We discuss this choice in the later section.
At the same time, we are interested in analyzing the situation, when the observers actually do observe the modes as closely as it is allowed by fundamental limits.
Thus, we will take $\phii$ to be localized in the same region as $\psii$ and likewise $\phiii$, in the same region as $\psiii$.
The other restriction is the orthonormality of the modes, which can be expressed as
\begin{equation}
\label{nooverlap}
(\phi_{\text{I}},\phi_{\text{II}}^{(\star)})=(\psi_{\text{I}}, \psi_{\text{II}}^{(\star)})=(\phi_{\text{I}}, \psi_{\text{II}}^{(\star)})=(\phi_{\text{II}},\psi_{\text{I}}^{(\star)})=0,
\end{equation}
where the symbol ${}^{(\star)}$ means that the above holds both with and without complex conjugation.
From this, for the ladder operators it follows that:
\begin{equation}
\left[\hat{f}_{\text{I}},\hat{f}_{\text{II}}^{(\dag)}\right] = \left[\hat{d}_{\text{I}},\hat{d}_{\text{II}}^{(\dag)}\right] = \left[\hat{f}_{\text{I}},\hat{d}_{\text{II}}^{(\dag)}\right] = \left[\hat{f}_{\text{II}},\hat{d}_{\text{I}}^{(\dag)}\right] =0.
\end{equation}

The construction, restrictions, and general discussion of the choice of such field decompositions are presented in Ref.~\cite{Ahmadi2016a} in an educational manner.
All the assumptions and properties demonstrated there stay the same also for higher dimensional cases.

\subsection{Modified Rindler coordinates and the field decomposition}
\begin{figure}[ht]
	\begin{center}
		\includegraphics[width=0.8\linewidth]{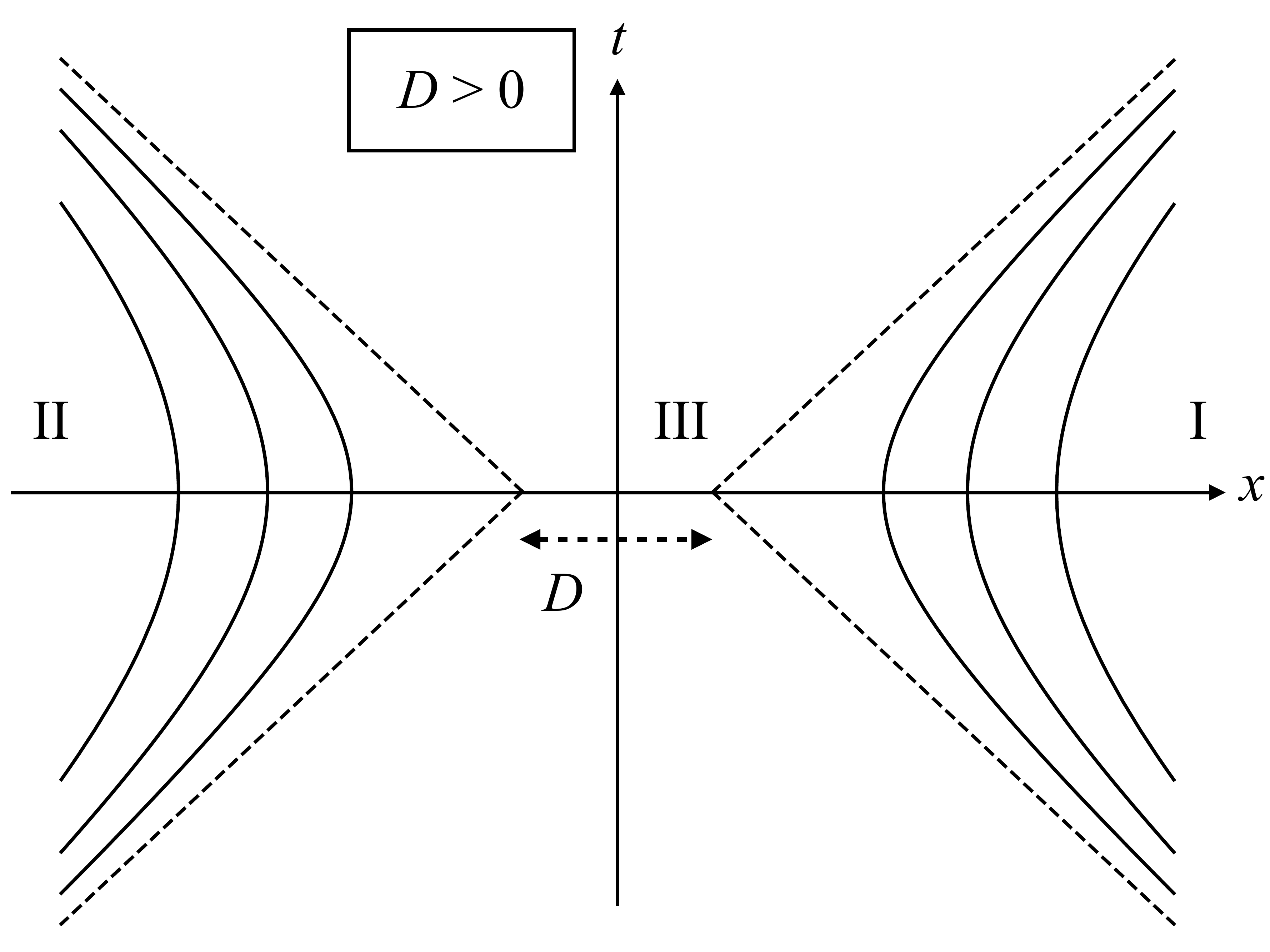}
		\includegraphics[width=0.8\linewidth]{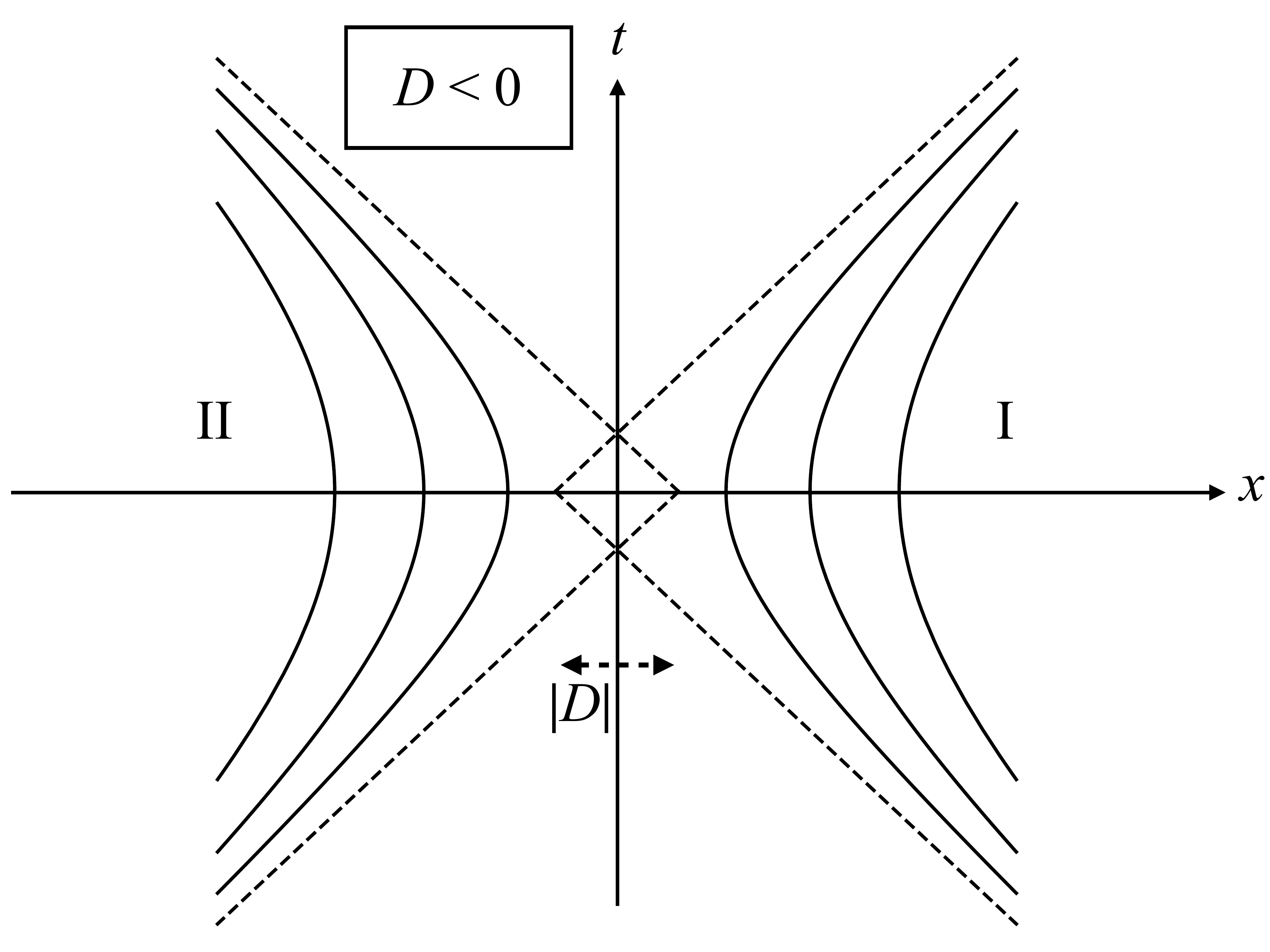}
		\includegraphics[width=0.8\linewidth]{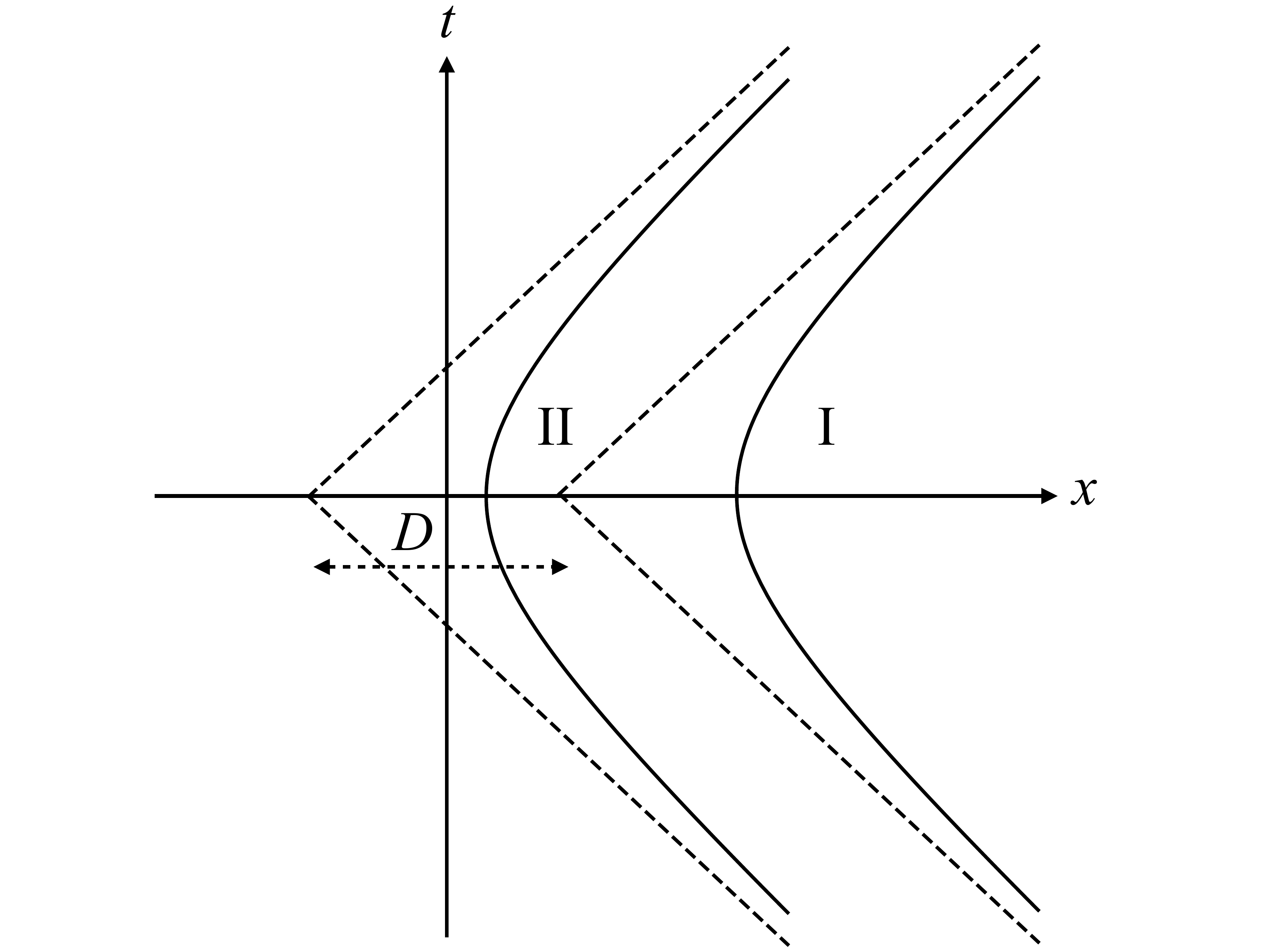}
		\caption{\label{schemes} (Top) When Rindler wedges I and II do not have a common
			apex and the two regions do not overlap.
			(Center) When Rindler wedges I and II do not have a common
			apex and the two regions overlap.
			(Bottom) Parallel accelerations with an additional distance $D$.
			Two wedges do not posses a common apex and both observers accelerate in the same direction.
			Figure from Ref.~\cite{Ahmadi2016a}.
		}
	\end{center}
\end{figure}

In order to analyze an accelerated frame of reference, we employ coordinates that provide a natural description of it -- the Rindler coordinates, $\chi$ and $\eta$~\cite{Wald1984,Takagi1986,Crispino2008}.
Our aim is to describe two accelerating observers, moving along $z$-axis, that are separated by the distance $D$ at their closest approach.
By the introduction of the modified Rindler coordinates~\cite{Ahmadi2016a}:
\begin{align} \label{opus_b3_mod_Rindler}
t &= \chi\sinh a\eta, \nonumber
\\
x &= x, \nonumber
\\
y &= y, \nonumber
\\
z &= \chi\cosh a\eta \pm \frac{D}{2},
\end{align}
we can freely tune that distance and consider separation between observers and their accelerations independently.
We will identify a constant acceleration $\mathcal{A}$ of a localized observer with their position in a Rindler chart, $\chi=1/\A$.
The upper sign in~\eqref{opus_b3_mod_Rindler} corresponds to the coordinates covering region I ($z>|t|+D/2$) for which $\chi>0$, and the lower sign corresponds to the region II ($z<|t|+D/2$) for which $\chi<0$.
In contrast to the standard coordinates, two wedges do not necessarily have a common apex at the origin -- it occurs only when $D=0$.
The additional separation can be either positive or negative, as in Ref.~\cite{Ahmadi2016a} (see Fig.~\ref{schemes}).
The case $D>0$ accounts for the situation in which two wedges are separated and additional region III has to be introduced for the field $\Phi$ to be completely specified on a Cauchy surface.
When $D$ is negative, two regions partially overlap what causes overcompleteness of the basis spanned by Rindler modes from the individual wedges.
The original Rindler coordinates are retrieved for $D=0$.
Moreover, we distinguish between counter-accelerated and co-accelerated cases.
The former (see top and center of Fig.~\ref{schemes}) is characterized by two observers accelerating in opposite directions, while in the latter, the observers accelerate in the same direction (see bottom of Fig.~\ref{schemes}).
Furthermore, we note that parameter $a$ should not be confused with the proper acceleration along a uniformly accelerated trajectory, and similarly $\eta$ should not be confused with the proper time along such a trajectory.

We will now proceed to decompose the quantum field $\Phi$ with respect to timelike Killing vectors associated with given, inertial and accelerating, sets of coordinates.
Firstly, in the Minkowski coordinates, the Klein-Gordon equation yields the following plane wave mode solutions:
\begin{equation}\label{u}
u_{\bf k}=
\frac{1}
{\sqrt{(2\pi)^32\sqrt{k^2+m^2}}}
e^{-i\sqrt{k^2+m^2}\,\,t+i{\bf k}\cdot{\bf x}}.
\end{equation}
Defining the corresponding annihilation operators $a_{\bf k}$, one can decompose the field $\hat{\Phi}$ into these modes as:
\begin{equation}
\hat{\Phi}=\int\text{d}^3{\bf k}
\left(
\hat{a}_{\bf k}u_{\bf k}+
\hat{a}\d_{\bf k}u_{\bf k}\s
\right).
\end{equation}
In the above formulae $\bf k$ is a 3-dimensional wave-vector.
This index may later be replaced with ${k_z\bf k_\perp}$, which is merely writing separately its components parallel and perpendicular to the z-axis.

When the Klein-Gordon equation is expressed in the Rindler coordinates (unmodified), it yields the following mode solutions~\cite{Crispino2008}:
\begin{align} \label{w}
\wi[\Omega{\bf k_\perp}]=&
\sqrt{\frac{\sinh\left(\frac{\pi\Omega}{a}\right)}{4\pi^4a}}
K_{\frac{i\Omega}{a}}\left(\sqrt{k_\perp^2+m^2}\,\,\chi\right)
e^{i{\bf k_\perp}\cdot{\bf x_\perp}-i\Omega\eta},
\\
\wii[\Omega{\bf k_\perp}]=&
\sqrt{\frac{\sinh\left(\frac{\pi\Omega}{a}\right)}{4\pi^4a}}
K_{\frac{i\Omega}{a}}\left(-\sqrt{k_\perp^2+m^2}\,\,\chi\right)
e^{i{\bf k_\perp}\cdot{\bf x_\perp}+i\Omega\eta},
\end{align}
where it is understood that the former ones span wedge I and are zero elsewhere, and the latter ones span wedge II and are zero elsewhere.
Here $\Omega$ is a positive parameter, called Rindler frequency; the two-dimensional vector ${\bf k_\perp}$ is the component of the wave-vector perpendicular to the $z$-axis. 

The mode decomposition of the field $\hat{\Phi}$ reads:    
\begin{align}\label{opus_dec_b_3d}
\hat{\Phi}=\int_0^{\infty}&\text{d}\Omega\int\text{d}^2{\bf k_\perp}
\big( 
\wiOk \biOk + \wiOk\s \biOk\d  \nonumber
\\
+ &\wiiOk \biiOk + \wiiOk\s \biiOk\d
\big)
+
\hat{\Phi}_\text{III}(D).
\end{align}
When $D=0$, Rindler modes~\eqref{w} constitute a complete basis for the field operator.
However, when $D \neq 0$, the additional part of the field decomposition, $\hat{\Phi}_\text{III}(D)$ has to be introduced in order to make a decomposition complete when $D>0$ or to prevent the basis from being overcomplete, when $D<0$.
We have to keep in mind that for every $D$ the mode decomposition is different, as both the Rindler modes are appropriately shifted -- $\psii$ and $\wi[\Omega{\bf k_\perp}]$ undergo a shift by $+\frac{D}{2}$ along the $z$-axis, and $\psiii$ and $\wii[\Omega{\bf k_\perp}]$ undergo a shift by $-\frac{D}{2}$ along the $z$-axis.
Every time we invoke scenario in which $D$ is nonzero, we will mean the use of such shifted wavepackets and modes.
Of course according Bogolyubov transformation between these shifted Rindler modes and Minkowski modes will be different, but we keep it in mind and make appropriate alternations.
As long as our wavepackets do not extend into region III, we are not interested in a specific form of $\hat{\Phi}_\text{III}$ as it virtually does not matter, not showing up in the calculations.

Moreover, the necessary conditions for only positive frequency contribution of our wave packets take form:
\begin{align} \label{ddecb}
\hat{f}_\text{I}&=\int\text{d}^3{\bf k}\,\,
\left( \phii , u_{\bf k} \right)  \hat{a}_{\bf k}, \nonumber
\\
\hat{f}_\text{II}&=\int\text{d}^3{\bf k}\,\,
\left( \phiii , u_{\bf k} \right)  \hat{a}_{\bf k}, \nonumber
\\
\hat{d}_\text{I}&=\int\text{d}\Omega\int\text{d}^2{\bf k_\perp}\,\,
\oI \hat{b}_{\text{I}\Omega\bf k_\perp}, \nonumber
\\
\hat{d}_\text{II}&=\int\text{d}\Omega\int\text{d}^2{\bf k_\perp}\,\,
\oIIk \hat{b}_{\text{II}\Omega\bf k_\perp}.
\end{align}

\subsection{The Gaussian channel}
We now need to characterize the quantum channel that transforms the state between two bases.
At first, we need to perform a Bogolyubov transformation between the modes and then ignore all the modes except for $\psii$ and $\psiii$.
Both of these actions preserve Gaussianity of the state~\cite{Werner1989}, so the channel is indeed a Gaussian one, as is our choice of the state, motivated by this feature.
Any bosonic Gaussian state is fully described by the first and second moment of its quadrature operators.
Let us define the following vector of quadrature operators, for the state of the inertial modes:
\begin{align}\label{QO}
\hat{\vec{X}}^{(f)}=(\frac{\hat{f}_{\text{I}}+\hat{f}_{\text{I}}^{\dag}}{\sqrt{2}},\frac{\hat{f}_{\text{I}}-\hat{f}_{\text{I}}^{\dag}}{\sqrt{2}i},\frac{\hat{f}_{\text{II}}+\hat{f}_{\text{II}}^{\dag}}{\sqrt{2}},\frac{\hat{f}_{\text{II}}-\hat{f}_{\text{II}}^{\dag}}{\sqrt{2}i})^T.
\end{align}
Then the vector of first moments, $\vec{X}^{(f)}$, and the matrix of second moments, $\sigma^{(f)}$, known as a covariance matrix, are defined as: 
\begin{subequations}\label{CMFM}
	\begin{align}
	X^{(f)}_k & = \<\hat{X}_{k}^{(f)}\>,\label{FM}\\
	\sigma^{(f)}_{kl} &=\left\langle \left\{\hat{X}^{(f)}_{k}- X_{k}^{(f)},\hat{X}^{(f)}_{l}- X_l^{(f)}\right\}\right\rangle\label{CM},
	\end{align}
\end{subequations}
where the anti-commutator is $\{\hat{A},\hat{B}\}=\hat{A}\hat{B}+\hat{B}\hat{A}$.
We make the same definitions for the accelerated modes.
They are exactly the same, except that the superscripts ${}^{(f)}$ should be replaced with ${}^{(d)}$.

A Gaussian channel acting on a Gaussian state can be desrcibed fully in terms of the first and the second moments, as~\cite{Schumaker1986}:
\begin{subequations}\label{GC}
\begin{align}
\vec{X}^{(d)} &=  M \vec{X}^{(f)},\label{GCFM}\\
\sigma^{(d)} & =  M\sigma^{(f)} M^T+N,
\label{GCCM}
\end{align}
\end{subequations}
where $M$ and $N$ are real, positive-defined $4\times 4$ matrices, whose specification uniquely characterizes the channel.
$N$ is symmetric and called the noise matrix, while $M$ is symplectic~\cite{Simon1988}.

The reader may also note that the form of the channel is independent of any coordinate system.
This feature allows to study such diverse scenarios as described by the modified Rindler coordinates.
For simplicity, we further assume that besides the observation of the wavepackets taking place at hypersurface $t=0$, there is also no relative velocity between the observers.
It removes a potential Lorentz boost from the calculations, but still allows the trajectories of the observers to be at different angles.
Firstly, we focus on collinear case, but later we explore the possibility of skew-oriented observers.
The expressions for $M$ and $N$ will now be derived, for the cases and under the assumptions discussed in Ref.~\cite{Ahmadi2016a}.
\section{Computing the Gaussian quantum channel\label{compchannel}}
The computation of the matrix $M$ does not use any coordinate system as it relies only on the overlaps of the inertial and the accelerating modes.
This computation does not differ from the $1+1$-dimensional case and therefore the result can be readily found in Ref.~\cite{Ahmadi2016a}:
\begin{widetext}
\begin{equation}\label{M}
M=
\left(\begin{matrix}
\text{Re}(\ali-\bei) & -\text{Im}(\ali+\bei) & 0 & 0 \\
\text{Im}(\ali-\bei) & \text{Re}(\ali+\bei) & 0 & 0 \\
0 & 0 & \text{Re}(\alii-\beii) & -\text{Im}(\alii+\beii) \\
0 & 0 & \text{Im}(\alii-\beii) & \text{Re}(\alii+\beii) 
\end{matrix}\right).
\end{equation}
\end{widetext}
In this expression we have defined: $\ali=(\psii,\phii)$ and $\bei=-(\psii,\phii^\star)$, and analogously for quantities with subscript~II.
In particular, matrix $M$ is shared between all the scenarios, regardless of the dimensionality, distance $D$, or the skewness of the observers.
However, this is not the case for the noise matrix $N$, and we have to compute it independently for each considered case.

We choose to express the noise matrix elements with the help of the Rindler basis.
As the noise matrix is independent of the initial state, without loss of generality, we find the noise matrix from the formula \eqref{GCCM} with Minkowski vacuum as the input state, i.e. $\sigma^{(f)}_{\text{vac}}=\I$.
The result is:
\begin{equation}\label{N}
N = \sigma^{(d)}_{\text{vac}} - MM^T,
\end{equation}
where $\sigma^{(d)}_{\text{vac}}$ is the corresponding output state.
The resultant noise matrix can be expressed in a simple form (for a detailed calculation, see Appendix~\ref{computingnoise}):
\begin{align}
\label{noise}
N=&
\left(\begin{matrix}
1+N_{\text{I}}
&
0
&
\Re\,N_{\text{I,II}}^+
&
\Im\,N_{\text{I,II}}^-
\\
0
&
1+N_{\text{I}}
&
\Im\,N_{\text{I,II}}^+
&
-\Re\,N_{\text{I,II}}^-
\\
\Re\,N_{\text{I,II}}^+
&
\Im\,N_{\text{I,II}}^+
&
1+N_{\text{II}}
&
0
\\
\Im\,N_{\text{I,II}}^-
&
-\Re\,N_{\text{I,II}}^-
&
0
&
1+N_{\text{II}}
\end{matrix}\right)-M M^T,
\end{align}
where, independently of $D$, the diagonal terms equal:
\begin{align}
N_\Lambda \label{N1D0}
=&
\int\text{d}\Omega\int\text{d}^2{\bf k_\perp}
\frac{|(\psi_\Lambda, w_{\Lambda\Omega{\bf k_\perp}})|^2}{\sinh\left(\frac{\pi\Omega}{a}\right)}
e^{-\frac{\pi\Omega}{a}},
\end{align}
and the off-diagonal terms
\begin{align}  
\label{opus_N12general}
N_\text{I,II}^\pm &\equiv 2
{}_M\bra{0} \hat{d}_\text{I} \hat{d}_\text{II} \pm \hat{d}_\text{I} \hat{d}_\text{II}\d \ket{0}_M.
\end{align}
are to be calculated in each case separately.
Here, $\Lambda$ labels each observer's wedge, $\Lambda=\{\text{I},\text{II}\}$
The diagonal terms are interpreted as the thermal noise due to the Unruh effect.

We first focus on the counter-accelerated case -- both observers accelerate in opposite directions and reside in their respective wedges.
Despite the directions of their accelerations are opposite, the magnitudes can be tuned independently.
In this setup we keep $\phii$ localized in wedge I, and $\phiii$ in wedge II of the Rindler chart.
When $D=0$, we obtain $N_\text{I,II}^+=N_\text{I,II}^-$ and:
\begin{equation} \label{N12D0}
N_\text{I,II}^\pm
=
\int\text{d}\Omega\int\text{d}^2{\bf k_\perp}
\frac{\oI\oIImk}{\sinh\left(\frac{\pi\Omega}{a}\right)}.
\end{equation}

For $D\neq 0$ the result is more complicated and presented in the Appendix~\ref{offdiag}.

Let us now move on to the parallel-accelerated case.
In this setup the two Rindler wedges are again shifted by $\pm\frac{D}{2}$, but the one on the left is now flipped, such that both observers accelerate towards the direction of increasing $z$.
We label the one on the right by I and the one on the left by II.
We proceed with the calculation, making sure that the condition \eqref{nooverlap} is satisfied.
It should be noted that this is possible even when $D=0$, but the modes need to be localized at different position, and hence the accelerating observers need to have different proper accelerations.
The details of the calculation are shown in the Appendix~\ref{Dnon0}.
The diagonal elements are the same as in the counter-accelerated case.
The off-diagonal blocks are different, and the result for $D=0$ is:
\begin{equation} \label{N12D0par}
N_\text{I,II}^\pm
=\pm
\int\text{d}\Omega\int\text{d}^2{\bf k_\perp}
\frac{\oI\oIIk\s}{\sinh\left(\frac{\pi\Omega}{a}\right)}
e^{\frac{\pi\Omega}{a}},
\end{equation}
The result for $D \neq 0$ is presented in the Appendix~\ref{offdiag}.

\section{The choice of modes\label{modechoice}}
Apart from some small remarks, we have not specified how to properly choose the input and the output modes.
The choice strictly follows the one from $1+1$-dimensional case, with a Gaussian envelope and sinusoidal modulation in perpendicular direction:
\begin{widetext}
\begin{align} \label{modephi}
&\phi_{\Lambda}\Big|_{t=0}=
\mathcal{N}_\phi\,\,
e^{-2(\frac{1}{\A_\Lambda L_\pa}\ln(\A_\Lambda z))^2-\frac{2}{L_\pe^2}(x^2+y^2)}
\sin\left[\sqrt{\Omega_0^2-m^2}\left(z\mp\frac{1}{\A_\Lambda}\right)\right]
\sin[\kappa_\pe x]\sin[\kappa_\pe y], \nn \\
&\psi_{\Lambda}\Big|_{t=0}=
\mathcal{N}_\psi\,\,
e^{-2(\frac{1}{\A_\Lambda L_\pa}\ln(\A_\Lambda\chi))^2-\frac{2}{L_\pe^2}(x^2+y^2)}
\Im\left\{
I_{-\frac{i\Omega_0}{\A_\Lambda}}\left(\frac{m}{\A_\Lambda}\right)
I_{\frac{i\Omega_0}{\A_\Lambda}}\left(m\,\chi\right)
\right\} 
\sin[\kappa_\pe x]\sin[\kappa_\pe y], \nn \\
&\text{with}\quad\quad 
\partial_t\phi_{\Lambda}\Big|_{t=0}=
-i\Omega_0\phi_{\Lambda}\Big|_{t=0}
\quad\quad\text{and}\quad\quad
\partial_\tau\psi_{\Lambda}\Big|_{\tau=0}=
\mp i\Omega_0\psi_{\Lambda}\Big|_{\tau=0},
\end{align}
\end{widetext}
where the upper sign refers to $\Lambda=\text{I}$ and the lower one to $\Lambda=\text{II}$, $\frac{1}{\A_\Lambda}$ is the position around which the mode function is centered, $L_\perp$ and $L_\parallel$ are perpendicular and parallel widths of the wavepackets.
Furthermore, $\kappa_\pe$ is the wave-vector in the perpendicular direction.
The normalization factors $\N_\phi$ and $\N_\psi$ have to be evaluated numerically.
The frequency $\Omega_0$, about which the spectrum is centered, has to be sufficiently large to effectively damp the negative frequencies, i.e. $\Omega_0 \gg 1/L_\parallel$.
Additionally, we introduce numerical positive-frequency cut-off for both wavepackets
\begin{align}
\phi_\Lambda & \to  \frac{\int\text{d}^3{\bf k}\,\, \left( u_{\bf k}, \phi_\Lambda \right) u_{\bf k} }{\sqrt{\int\text{d}^3{\bf k}\,\, \left|\left( u_{\bf k}, \phi_\Lambda \right) \right|^2  }} \\
\psi_\Lambda& \to  \frac{\int\text{d}\Omega\int\text{d}^2{\bf k_\perp}\, \oIokL w_{\Lambda\Omega{\bf k_\perp}} }{\sqrt{\int\text{d}\Omega\int\text{d}^2{\bf k_\perp}\, \left|\oIokL \right|^2  }}.
\end{align}
However, our choice of the analytical form of the mode functions renders them almost changeless after this operation.

\begin{figure}[ht]
	\begin{center}
		\includegraphics[width=1.0\linewidth]{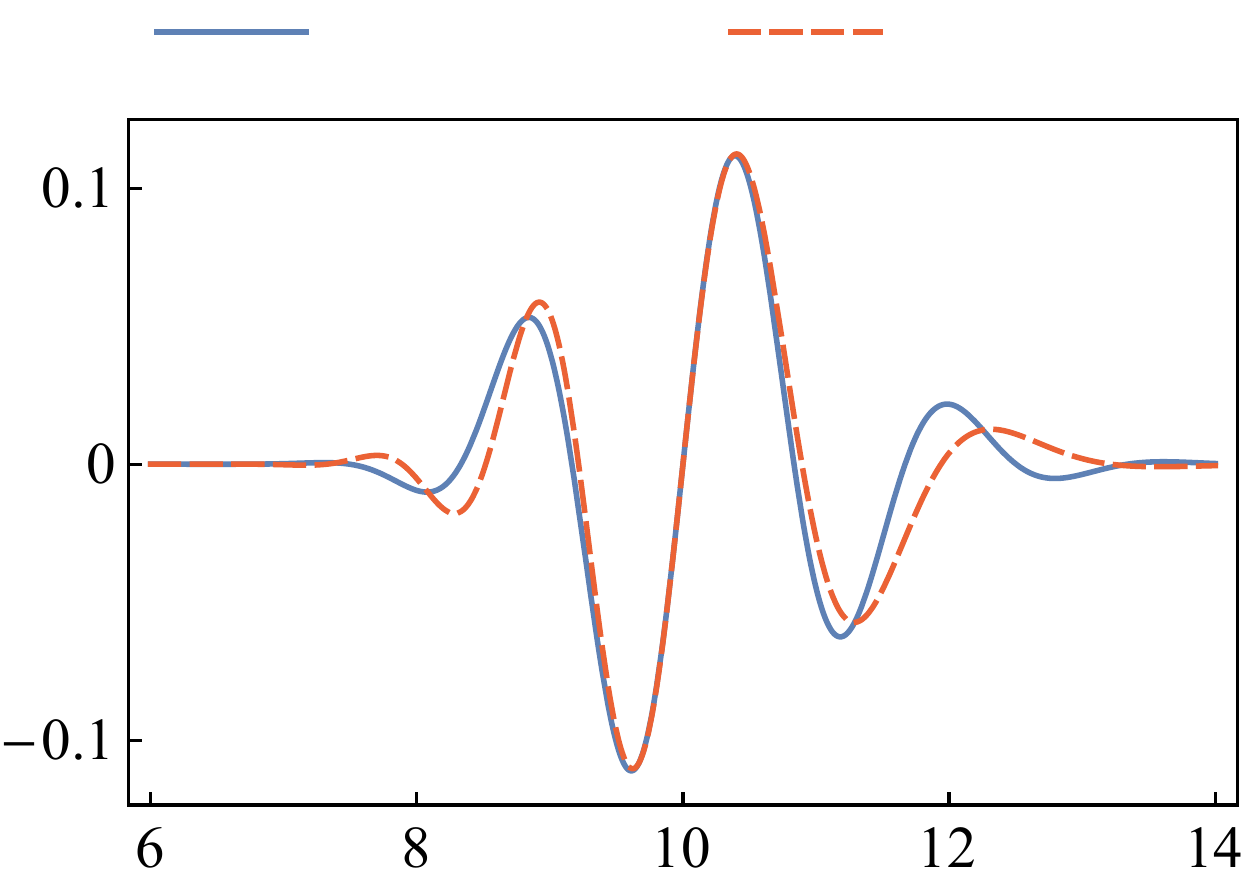}
		\caption{\label{modes1} Comparison of the spatial dependence in $z$-direction for the modes $\phi$ and $\psi$ for the following choice of parameters: $\A=0.1$, $L_{||}=L_\pe=2$, $\Omega_0=4.71$, $m=0.1$, $\kappa_\pe=2$, along the $z$-axis at $x=y=1$. }
	\end{center}
\end{figure}

Finally, in Fig.~\ref{modes1} we plot the comparison of the spatial dependence in $z$-direction for the input and output modes for one set of parameters.
The mode mismatch that causes the degradation of the entanglement is clearly seen.
We can now proceed to the qualitative considerations of the vacuum entanglement.
\section{Entanglement of the vacuum\label{vacent}}
We proceed to present the results for the case when the input state of the channel is the Minkowski vacuum state, i.e. $\vec{X}^{(f)}=\vec{0}$ and $\sigma^{(f)} = \openone$.
From \eqref{GC} and \eqref{N} we obtain the following expressions for the output state:
\begin{align}
\label{vacuumout}
\vec{X}^{(d)} &=  0,\nonumber \\
\sigma^{(d)} & =
\left(\begin{matrix}
1+N_{\text{I}}
&
0
&
\Re \,N^+_{\text{I,II}}
&
\Im\,N^-_{\text{I,II}}
\\
0
&
1+N_{\text{I}}
&
\Im\,N^+_{\text{I,II}}
&
-\Re\,N^-_{\text{I,II}}
\\
\Re\,N^+_{\text{I,II}}
&
\Im\,N^+_{\text{I,II}}
&
1+N_{\text{II}}
&
0
\\
\Im\,N^-_{\text{I,II}}
&
-\Re\,N^-_{\text{I,II}}
&
0
&
1+N_{\text{II}}
\end{matrix}\right).
\end{align}

To quantify the amount of entanglement in the above state, we choose to use the logarithmic negativity,
 ${\cal E_N}$.
This quantity is a measure of distillable entanglement and is particularly easy to compute for any two-mode Gaussian state~\cite{Adesso2007a}.
For the output state $\sigma^{(d)}$ given by Eq.~\eqref{vacuumout} the logarithmic negativity is equal to:
\begin{align}\label{logneg}
{\cal E_N} = \max\left\{0,-\log\sqrt{\frac{\Delta-\sqrt{\Delta^2-4\det \sigma^{(d)}}}{2}}\right\},
\end{align}
where $\Delta\equiv (1+N_{\text{I}})^2 +  (1+N_{\text{II}})^2 + 2 \Re N^+_{\text{I,II}} \Re N^-_{\text{I,II}} + 2 \Im N^+_{\text{I,II}} \Im N^-_{\text{I,II}}$.
In the lowest order it simplifies to
\begin{align}\label{logneg2}
{\cal E_N} = \max\Bigg\{0,\frac{1}{2}\Bigg( & \sqrt{(N_{\text{I}}-N_{\text{II}})^2 +\left|N^+_{\text{I,II}}+N^-_{\text{I,II}}\right|^2} \nonumber \\
&-N_{\text{I}}-N_{\text{II}} \Bigg) \Bigg\}.
\end{align}

Note that the expression for the logarithmic negativity in this particular case has no dependence on the overlaps of the inertial and the accelerating mode functions.
This is a special exception that holds only for a coherent state -- for other two-mode states, e.g. the squeezed states, the entanglement degradation due to the mode mismatch is the dominating effect~\cite{Grochowski2017}.
As for the Minkowski vacuum state, the logarithmic negativity is zero from the beginning, so the Unruh noise can introduce some entanglement and quantum correlations.
Indeed, we show this to be the case.

We now proceed to evaluate integrals that express the noise matrix elements.
Additional dimensions in comparison to $1+1$-dimensional case~\cite{Ahmadi2016a} make the numerical computations more challenging.
Even in the simple case of $D=0$ they reduce to the triple integrals with highly oscillatory integrands.
For $D \neq 0$ the integrals become quintuple and as we are unable to efficiently evaluate them, we only stick to the $D=0$ case, which is still very time-consuming.
Let us focus on the counter-accelerated scenario, when $D=0$.
In this case the terms arising in the output state covariance matrix, are given by the formulae \eqref{N1D0}-\eqref{N12D0}.
When expressing the integrand in polar coordinates with respect to ${\bf k_\pe}$, the angular integral can be computed analytically.
This reduces the problem to a double integration, to be performed numerically.

The $3+1$-dimensional calculations confirm results for $1+1$-dimensional bosonic and fermionic fields~\cite{,Ahmadi2016a,Richter2017}.
Increasing either of the accelerations leads to more entanglement, which has a clear physical explanation.
It is well known that the Minkowski vacuum state, when viewed in the Rindler coordinates, becomes a tensor product of two-mode squeezed states of modes with the same momenta in wedges I and II~\cite{Crispino2008}.
The higher the acceleration, the higher temperature observers perceive and, as a consequence, the more correlated the Unruh noise gets.

Furthermore, entanglement is affected with the change of the longitudinal (parallel to the acceleration) size of the wavepacket $L_\parallel$ and with the change of the central frequency $\Omega_0$ consistently with Refs.~\cite{,Ahmadi2016a,Richter2017}.
The most entanglement is present, if the spectrum is more red-shifted, and if the width of the wavepacket is smaller.
This result is consistent with previous studies of the topic~\cite{Dragan2013b}.

\begin{figure}[ht]
	\centering
	\includegraphics[width=0.95\linewidth]{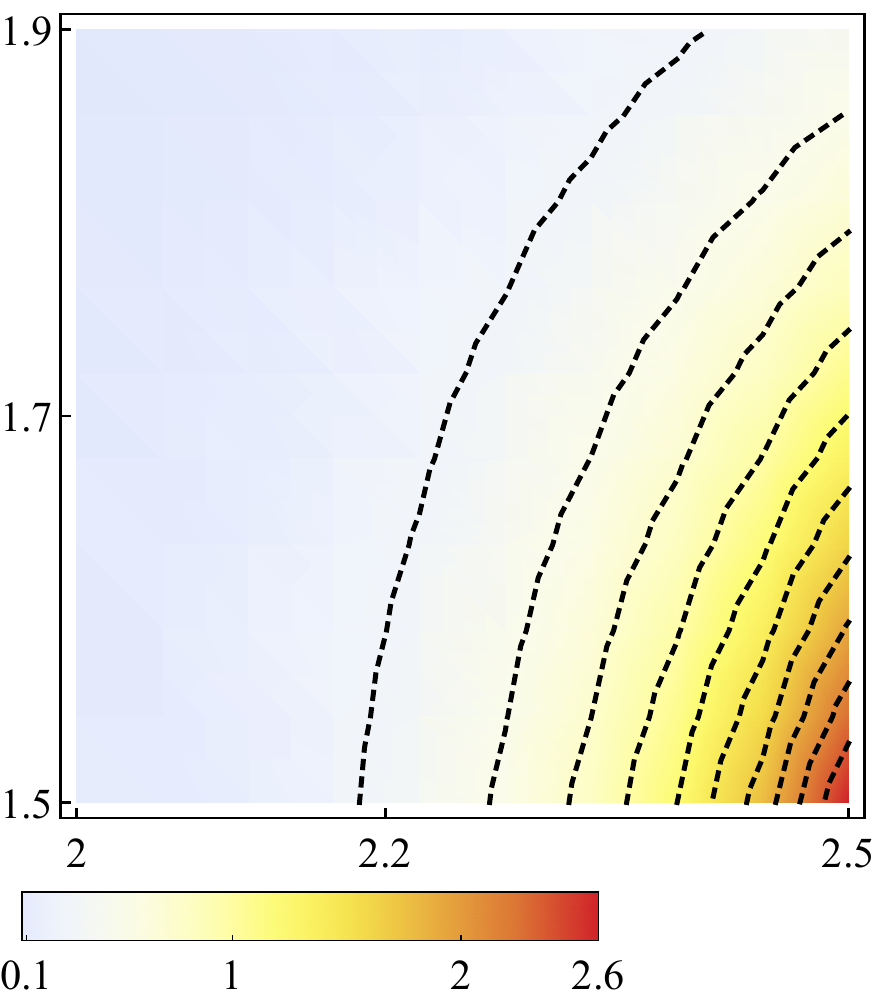}
	\caption{{Logarithmic negativity of the Minkowski vacuum for two counter-accelerated modes, as a function of $L_\pe$ and $\kappa_\pe$ for $D=0$.
			We have chosen $\A_\text{I}=\A_\text{II}=0.1$, $L_{||}=2$, $m=0.1$, $\Omega_0=4.71$. \label{fig_3d_plotac}}}
\end{figure}

To go beyond, we have also investigated the dependence on the parameters related to the dimensions perpendicular to the motion of the observers, namely on the width and wavenumber perpendicular to the $z$-axis.
The result is plotted in the Fig.~\ref{fig_3d_plotac}.
As expected, it is better to squeeze the wavepacket to get more entanglement, but contrary to a $\Omega_0$ decrease being beneficial, $\kappa_\pe$ needs to be increased to make logarithmic negativity larger.
As for our best knowledge, no such analysis has ever been performed, so we cannot compare our results with any previous ones.

Moving onto the co-accelerated case, for $D=0$ it is easily seen from Eq.~(\ref{N12D0par}) that $N_\text{I,II}^+=-N_\text{I,II}^-$, simplifying~(\ref{logneg2}) to
\begin{align}\label{logneg3}
{\cal E_N} \approx \max\left\{0,\frac{1}{2}\left( |N_{\text{I}}-N_{\text{II}}|-N_{\text{I}}-N_{\text{II}} \right) \right\}=0
\end{align}
for positive-valued $N_{\text{I}}$ and $N_{\text{II}}$, which is the case.
It means that in the lowest order there is no entanglement present.
Operationally, magnitudes of accelerations would have to be greatly larger in comparison to the counter-accelerated case to have any chance to detect entanglement.
This result holds for any dimensionality in such a setup.
Further numerical calculations in $1+1$-dimensional setting~\cite{Ahmadi2016a,Ahmadi2017} for $D \neq 0$ also found no entanglement, so it strongly suggests a similar result in higher dimensions.  

\section{Skew oriented accelerating observers}\label{r:2d}
In this Section we consider a setting in which two observers probe the Minkowski vacuum, but in contrast to previously investigated scenarios, their motion is not collinear.
The minimal framework in which we can describe such a situation is $2+1$-dimensional, so in order to suppress the computational difficulty we stick to such a system.
However, our framework can readily be used to study higher-dimensional problems.
\subsection{The setup}\label{skewsetup}
\begin{figure}[ht]
	\centering
	\includegraphics[width=0.9\linewidth]{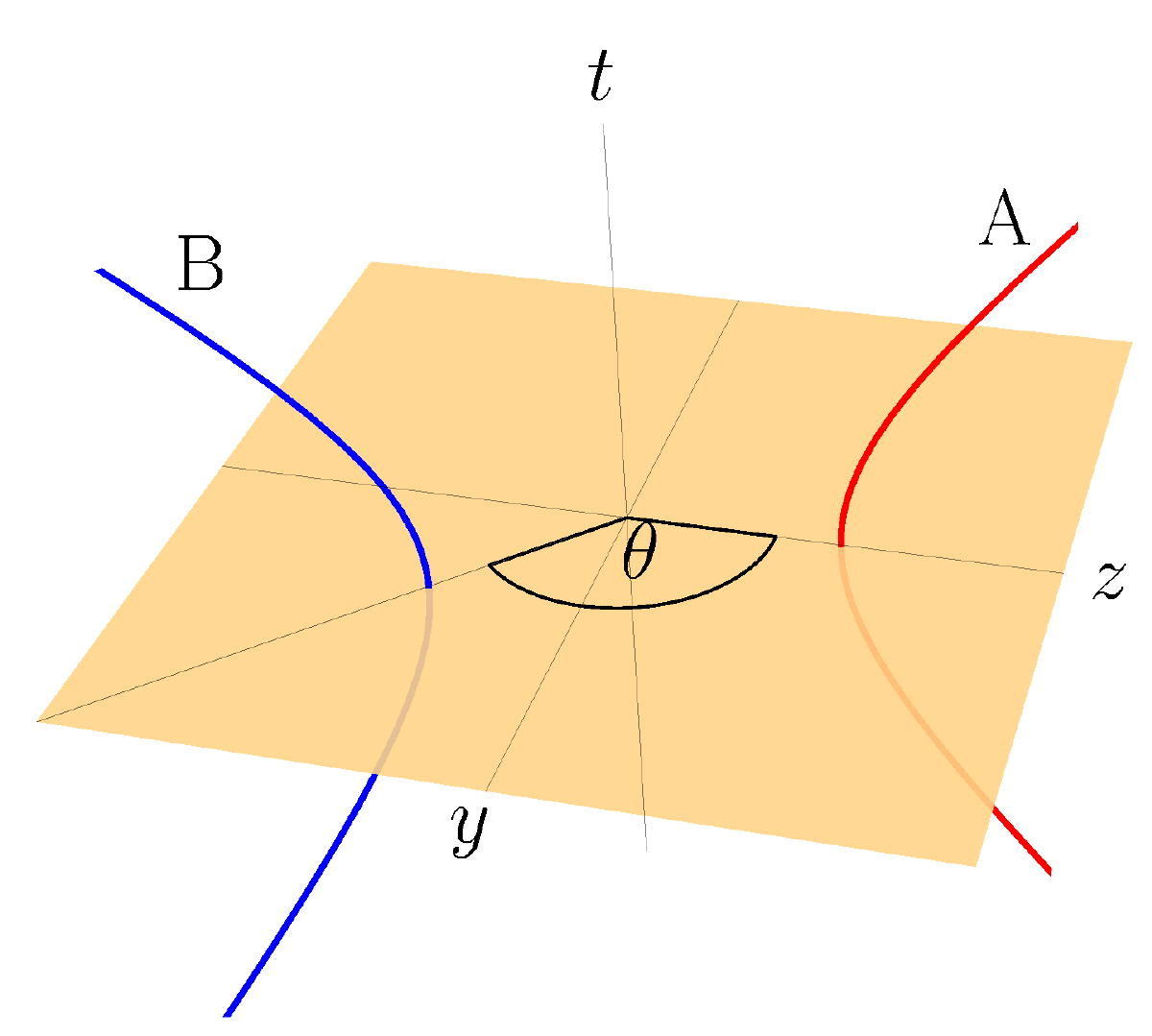}
	\caption{{The setup in which we study the vacuum entanglement witnessed by two non-collinearly moving observers, Alice and Bob.
			Alice's trajectory (red curve) is kept fixed, as she uniformly accelerates across $y=0$ plane, making measurement with her detector at $t=0$ and at the position $z=1/\A$.
			Bob's trajectory (blue curve) is positioned at relative angle $\theta$ with respect to Alice as he also uses his detector at $t=0$.
			$\theta=\pi$ and $\theta=0$ cases reproduce previously studied counter- and co-accelerated scenarios, respectively. 
			\label{scen1}}}
\end{figure}

Contrary to the previous Sections, we introduce names for the observers, Alice (A) and Bob (B), as their trajectories are no longer contained in Rindler wedges I and II.
In the scenario that we consider, trajectory of Alice stays fixed -- she uniformly accelerates along the $z$-axis with a proper acceleration $\A$.
As in usual situation, at $t=0$ she is at the position $z=\frac{1}{\A}$.
For the depiction of her trajectory, see Fig.~\ref{scen1}.
We choose Bob to also accelerate with a proper acceleration $\A$.
However, his trajectory is not collinear with respect to Alice and it is parametrized by the relative angle between the observers' directions of motion, $\theta$.
Moreover, we assume that both observers have access to the wavepackets that have the same spatial shape in their respective co-moving frames of reference.
The inertial wavepackets are prepared to match the orientation of both accelerating wavepackets.

At $t=0$ Bob is always at a distance $\frac{1}{\A}$ from the origin of the Minkowski frame.
Effectively, his trajectory is rotated around this origin by the angle $\theta$.
By putting $\theta=0$ we retrieve co-accelerated scenario from the last Section and in $\theta=\pi$ we end up at the counter-accelerated case.
Of course as we have chosen the magnitudes of the accelerations to be the same, by this kind of rotation we cannot achieve $\theta=0$ case as both wavepackets start to overlap.

\subsection{Characterization of the quantum channel}
The form of the channel~\eqref{GC} describing the effect of the acceleration remains unchanged.
Both the $M$ matrix and the diagonal parts of the noise matrix are computed in the same way as in $3+1$-dimensional unrotated case, with the only difference originating from the reduced dimensionality:
\begin{align}
N_\text{A} = N_\text{B} \label{N1D0s}
=&
\int\text{d}\Omega\int\text{d}{ k}
\frac{|(\psi, w_{\text{I}\Omega{ k}})|^2}{\sinh\left(\frac{\pi\Omega}{a}\right)}
e^{-\frac{\pi\Omega}{a}},
\end{align}
where we have already introduced a change of naming convention, $\text{I} \rightarrow \text{A}$ and $\text{II} \rightarrow \text{B}$.
In this Section, I and II will denote the right and left Rindler wedges with respect to the Alice's frame of reference.
This change of convention is caused by the way we evaluate the other elements of the noise matrix.

Because of our choice of identical wavepackets for Alice and Bob, we can see in Eq.~(\ref{N1D0s}) that the diagonal elements are equal, $N_\text{A} = N_\text{B}$.
Furthermore, by $\psi$ we have denoted the common spatial shape of the observers' wavepackets.
Note that for the calculation of $N_\Gamma$, where $\Gamma=\{\text{A},\text{B}\}$, no rotation is involved yet:
\begin{align}
(\psi_\Gamma, w_{\text{I}\Omega{ k}})=(\psi (y,z), w_{\text{I}\Omega{ k}}(y,z)).
\end{align}
Also, with~(\ref{N1D0s}) we have assumed that the wavepackets do not have any negative frequency contributions from the perspectives of their co-moving frames of reference.

We now proceed to the evaluation of the off-diagonal terms of the noise matrix.
We start by introducing additional two sets of coordinates -- rotated Minkowski frame:
\begin{align}\label{mprim}
t^{'}&= t,\nonumber \\
y^{'}&= (y-D_y) \cos \theta - (z-D_z) \sin \theta,\nonumber  \\
z^{'}&= (z-D_z) \cos \theta + (y-D_y) \sin \theta
\end{align}
and rotated Rindler frame:
\begin{align}\label{rindler}
t^{'}&= \chi^{'}  \sinh (a \eta^{'}),\nonumber \\
y^{'}&= y^{'},\nonumber  \\
z^{'}&= \chi^{'}  \cosh{ (a \eta^{'})},
\end{align}
where $D_y$ and $D_z$ are shifts along $y$ and $z$ axes, respectively.
This shift, along with a rotation, allows an arbitrary placement of Bob's trajectory, and as a result, analysis of more complicated geometries than the one considered in this Section.
We keep such a shift for the sake of generality.
Standard Minkowski frame is retrieved when $\theta=0$ and rotated Rindler chart is interpreted as the Bob's frame of reference.
In the following calculations, we will characterize the channel in terms of scalar products that will be evaluated in standard Minkowski coordinates.
It is also helpful for later brevity of notation to introduce the following shorthands for these scalar products:
\begin{align} \label{nota}
\Phi_{\Gamma \Lambda } (k)&\equiv\left( \psi_\Gamma, w_{\Lambda \Omega k} \right), \nonumber \\
\overline{\Phi_{\Gamma \Lambda}} (k)&\equiv\left( \psi_\Gamma, {w_{\Lambda \Omega k}^*} \right).
\end{align}
Here, the bar denotes an overlap with a negative-frequency Rindler mode, the first index, $\Gamma=\{\text{A},\text{B}\}$ -- which wavepacket is under consideration, and the second index, $\Lambda=\{\text{I},\text{II}\}$ -- which Rindler wedge is taken.
We skip the $\Omega$-dependence to further shorten the notation.
\begin{figure}[ht]
	\centering
	\includegraphics[width=0.8\linewidth]{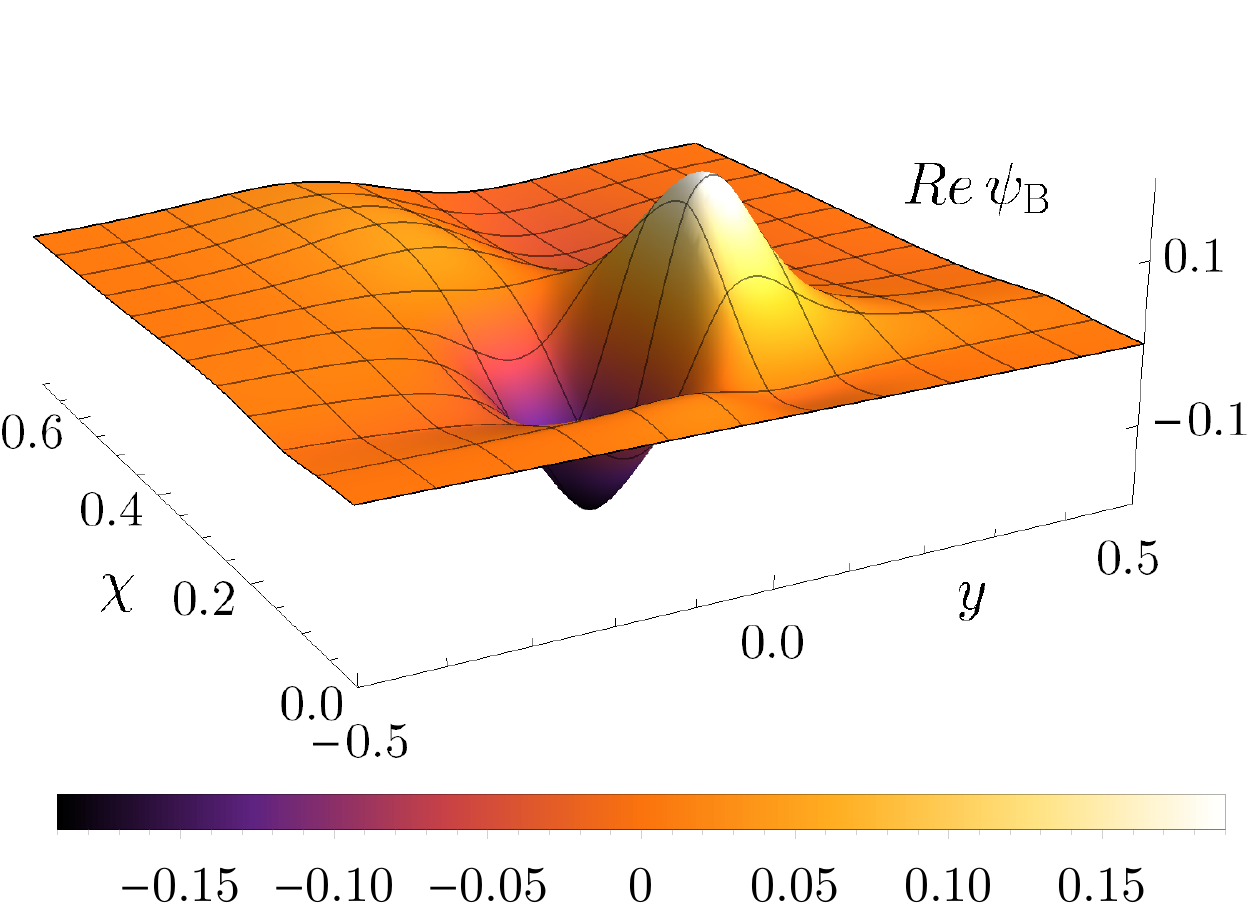}
	\includegraphics[width=0.8\linewidth]{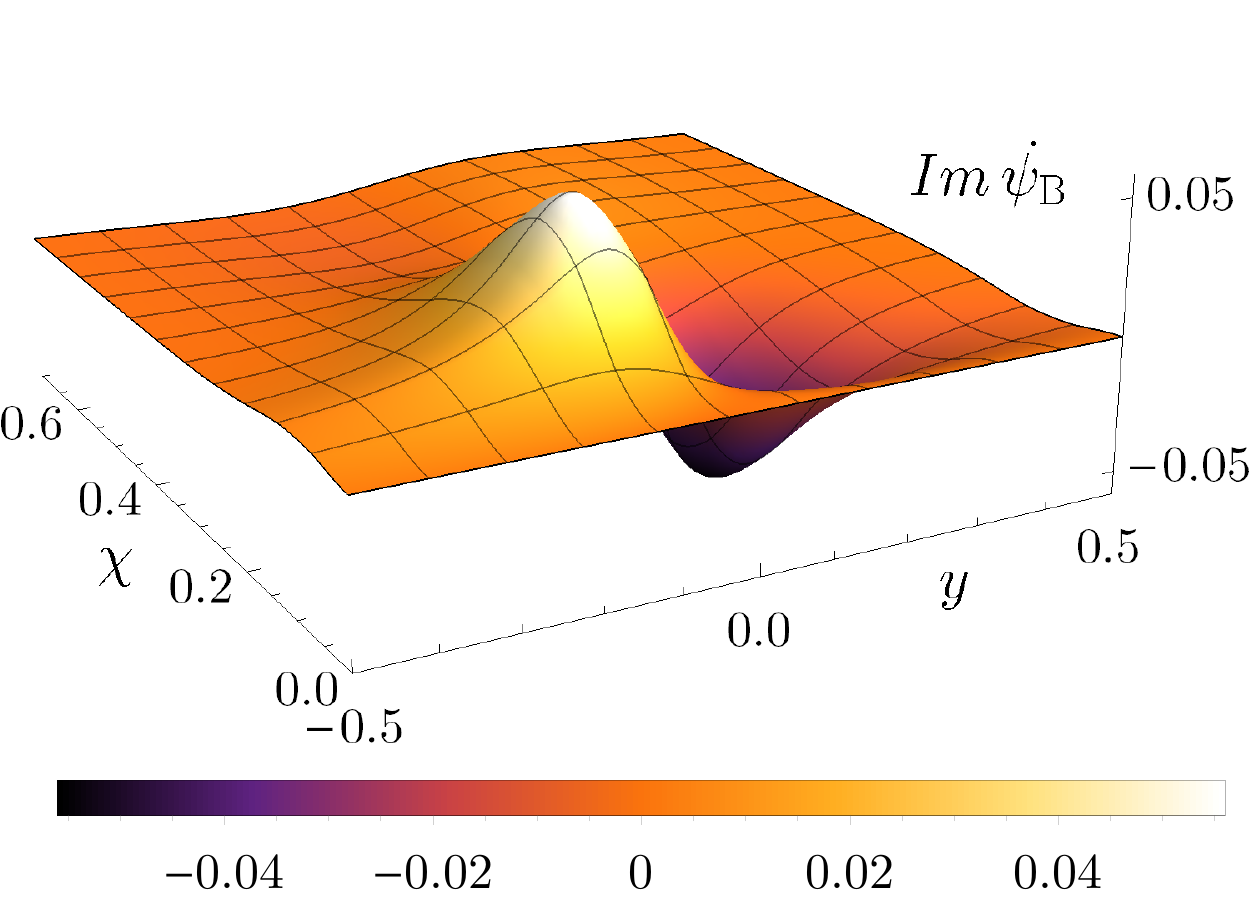}
	\caption{{The wavepacket used in the skew observers setup.
			In the top its spatial profile and in the bottom -- its time derivative.   \label{wv}}}
\end{figure}

Let us recall the form of the off-diagonal element of the noise matrix~(\ref{opus_N12general}):
\begin{align} \label{ns}
N_{\text{A}\text{B}}^{\pm} = \pm 2  {}_M\bra{0} \hat{d}_\text{A} \hat{d}_\text{B} \pm \hat{d}_\text{A} \hat{d}_\text{B}^\dagger \ket{0}_M.
\end{align}
We assume that both wavepackets consist of only positive frequencies in their respective frames, so for Alice we can write down
\begin{align} \label{ds}
\hat{d}_\text{A} = \int\text{d}\Omega\int\text{d}k \ \left( \psi_\text{A}, w_{\text{I} \Omega k} \right) \hat{b}_{\text{I} \Omega k},
\end{align}
where by $\psi_\text{A}$ we denote Alice's wavepacket.
In case of Bob's rotated wavepacket, the decomposition into Rindler modes associated with the Alice's frame is not constrained by any condition.
In general, it will contain negative-frequency contribution in his frame of reference:
\begin{align} \label{d2s}
\hat{d}_\text{B} = \int\text{d}\Omega\int\text{d}k \ \bigg( & \Phi_{\text{BI}} \hat{b}_{\text{I} \Omega k} +  \overline{\Phi_{\text{BI}}} \hat{b}_{\text{I} \Omega k}^\dagger +\nonumber \\
 & \Phi_{\text{BII}} \hat{b}_{\text{II} \Omega k} + \overline{\Phi_{\text{BII}}} \hat{b}_{\text{II} \Omega k}^\dagger    \bigg). 
\end{align}

By inserting~(\ref{d2s}) and~(\ref{ds}) into~(\ref{ns}) we arrive at the expressions for the elements of the matrix $N$:
\begin{align} \label{ns12}
N_{AB}^{\pm} = \int & \text{d}\Omega\int\text{d}k \ \Phi_{\text{AI}}(k) \times \nonumber \\\Big[  & \frac{e^{\pi \Omega  /a}}{\sinh{\pi \Omega /a}} \left( \overline{\Phi_{\text{BI}}} (k) \pm  \Phi_{\text{BI}}^*(k)  \right) + \nonumber \\
&\frac{1}{\sinh{\pi \Omega /a}} \left( \Phi_{\text{BII}} (-k) \pm  \overline{\Phi_{\text{BI}}}^*(-k)  \right)  \Big].
\end{align}
Their numerical evaluation is very complicated and is addressed in detail in Appendix~\ref{computingnoise2}.

\subsection{The choice of modes}
We planned to follow the choice of the modes from the unrotated case, but due to the computational difficulties we had to prepare a wavepacket that makes numerical evaluation as easy as possible.
The problems stem from the need to compute the negative-frequency contribution of the rotated wavepacket.
In the unrotated case we circumvent this issue as our expressions do not include such a part from the beginning.
The positive sector of the wavepacket's spectrum is given by an analytical formula and the only effect negative frequencies have is the change of the normalization constant.

In the rotated case, the effect of negative contributions is very subtle.
As it was stated in Ref.~\cite{Ahmadi2016a}, no wavepacket of only positive-frequency spectrum can have a finite size.
It is clearly seen in the numerical calculations that the finite region in which the integrals are evaluated and the behavior of the wavepacket's tails always cause a nonzero, however small, negative-frequency contribution.
This little part is irrelevant in some quantities like the norm of the wavepacket, but as the integrands of Eqs.~\eqref{ns12} are localized in the infrared end of the spectrum, it is no longer negligible.
The wavepackets utilized in the previous Section were localized at $\sim \frac{\Omega_0}{\A}$ in the frequency space, so their values for small $\Omega$ were little and error-prone.

In order to have an operational possibility to control the effect of the negative-frequency contribution on the integrals~\eqref{ns12}, we had to prepare a wavepacket for which the spectrum is localized for small values of $\Omega$.
We have managed to find such a wavepacket semi-numerically, depicted in Fig.~\ref{wv}.
The parameters of such a wavepacket are borderline satisfying the conditions described in detail in Ref.~\cite{Ahmadi2016a}.
It is relatively close to the horizon, described by $\A \sim 4.0$ and quite wide, with $L_\pa \sim 0.25$.
The method to create it is given in Appendix~\ref{computingnoise2}.

As for the numerical calculations, routines were written in both C and Fortran languages, and specifically, we have used an algorithm by A. Gil \textit{et al.} to compute modified Bessel functions of imaginary order and positive argument~\cite{Gil2004}.
To perform fast Fourier transforms we used free software \textit{Fastest Fourier Transform in the West}~\cite{Frigo1999}.

\subsection{Entanglement of the vacuum}
We now compute the amount of the entanglement witnessed by Alice and Bob while probing the Minkowski vacuum state.
To quantify it, we again use the logarithmic negativity~(\ref{logneg}), that given $N_\text{A}=N_\text{B}$ in the first order reduces to
\begin{align}\label{logneg4}
{\cal E_N} = \max\left\{0,\frac{1}{2}\left( |N_{\text{AB}}^++N_{\text{AB}}^-|-2 N_\text{A}  \right) \right\}.
\end{align}
\begin{figure}[hbtp!]
	\centering
	\includegraphics[width=0.8\linewidth]{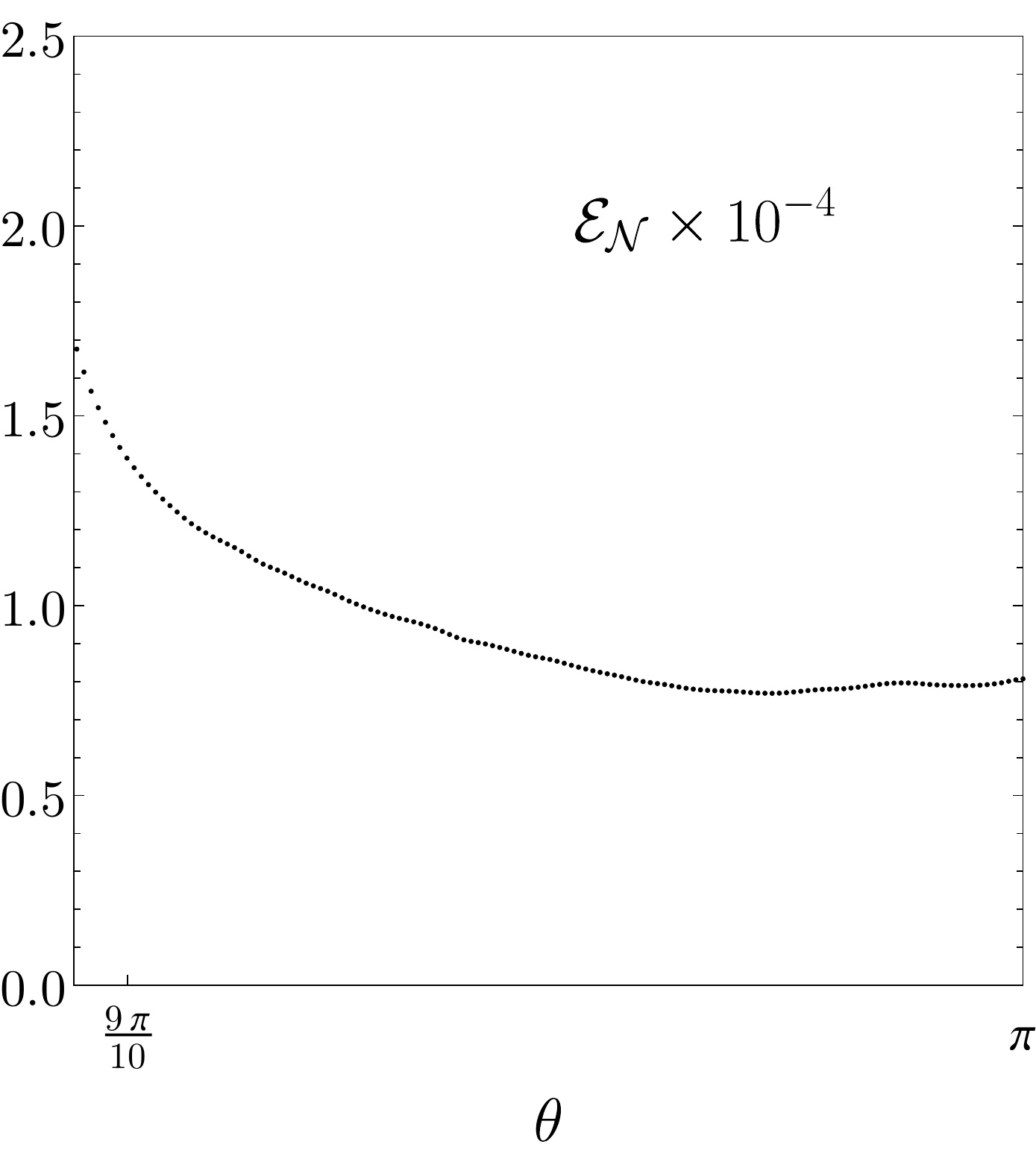}
	\caption{{Logarithmic negativity for a scenario involving non-collinear trajectories of the accelerated observers.
    Due to numerical difficulties, we restrict ourselves to the regime of a small rotation between the observers.
			A limit $\theta=\pi$ can be identified as a previously considered counter-accelerated case and a limit $\theta=0$ as a co-accelerated case.
			We can see that logarithmic entanglement grows as the relative angle between two counter-accelerating observers gets smaller.
			\label{lognegs}}}
\end{figure}

Due to numerical difficulties, we are able to control the integrals' convergence only for a small relative rotation of Alice and Bob.
In Fig.~\ref{lognegs} we plot the logarithmic negativity as a function of the relative angle $\theta$ from the interval $[0.9 \pi, \pi]$.
A limit $\theta=\pi$ can be identified as a previously considered counter-accelerated case and a limit $\theta=0$ as a co-accelerated one.
In this small interval, the logarithmic negativity grows while $\theta$ moves away from the counter-accelerating limit.
However, at the co-accelerating limit, we expect the entanglement to vanish, so our result suggests non-monotonic behavior of logarithmic negativity in between these two regimes.

\section{Conclusions\label{concl}}
In this work, we have studied how an arbitrary Gaussian state of two localized wavepackets of a massive real scalar field in $3+1$-dimensional Minkowski spacetime is described by a pair of uniformly accelerated observers.
Such a transformation can be formulated in terms of a noisy Gaussian channel, which can be expressed in a fully analytical way.
This channel has been used to study different scenarios of collinear motion of the observers.
In contrast to previous studies, this framework has allowed us to go beyond usual Rindler chart and to analyze an arbitrary relative rotation of the observers.
This way, we have derived the parameters of the channel for an arbitrary geometry.

Then, the entanglement of the Minkowski vacuum has been studied as seen by two such accelerating observers.
As expected, in the counter-accelerating case, we have observed that the vacuum entanglement is an increasing function of proper accelerations, when the two Rindler wedges have a common appex.
This enhancement has an operational meaning as the resulting entanglement can be extracted by a suitable pair of local detectors.
Also in this case, we have found out that in order to extract more entanglement, the width of the wavepackets has to be as small as possible in all the spatial directions.
This universal behavior is surprisingly not reproduced for the frequencies characterizing the wavepackets.
While it is beneficial to make the spectrum of the wavepacket more shifted to the red end in the direction of the motion, the opposite tendency can be observed for the perpendicular direction.

On the contrary, in the co-accelerating case, accelerations necessary to witness entanglement are much higher.
It renders this case operationally useless as compared to counter-accelerating scenario.
All of the findings for both cases stay in agreement with previous works on topic.

Moreover, we wanted to capitalize on the fact that we consider more dimensions than in $1+1$-dimensional case and to study a non-collinear motion of the observers.
In Section~\ref{r:2d} we have characterized the channel in such a setting and used it to analyze the vacuum entanglement.
Due to the numerical difficulties, we were able to obtain results only for a geometry that does not drastically differ from the collinear, counter-accelerating case.
However, we have showed that the entanglement behaves rather counter-intuitively, increasing while getting further from the counter-accelerating limit.

This work has touched the topic of relative skewness of two Rindler observers for the first time and has presented a framework that provided a constructive method to calculate observable quantities in such a setup.
Further effort into this direction can bear fruit in the form of geometries that could potentially make the measurement of the Unruh effect reachable by some form of enhancement of the correlations.
Moreover, the framework described here can be also readily applied to study any quantum information protocols that involve continuous variable systems and which can be affected by a gravitational force.
An analysis of other fields (fermionic, massless) in higher-dimensional spacetimes could also make an interesting development.
\acknowledgments
We thank Mehdi Ahmadi for useful discussions and comments. This work was supported by National Science Centre, Sonata BIS Grant No. DEC-2012/07/E/ST2/01402.

{\onecolumngrid

\appendix
\section{Computing the noise matrix $N$ in collinear case\label{computingnoise}}

In this appendix we show the details of the calculation of the noise matrix $N$.
In order to do this, we compute the output state of the channel for the Minkowski vacuum as the input state.
Then, using Eq.~\eqref{N} we can compute the noise matrix for $D=0$ and $D\neq 0$ cases.

\subsection{When Rindler wedges have a common apex ($D=0$)}\label{D0}
We first need the expectation values of all products of two Rindler ladder operators~\cite{Crispino2008}:
\begin{align}
{}_\text{M}\bra{0}\bi[\Omega{\bf k_\perp}] \bi[\Xi{\bf l_\perp}] \ket{0}_\text{M}
=&
0,
\\
{}_\text{M}\bra{0}\bi[\Omega{\bf k_\perp}]\d \bi[\Xi{\bf l_\perp}] \ket{0}_\text{M}
=&
\frac{e^{-\frac{\pi\Omega}{a}}}{2\sinh\left(\frac{\pi\Omega}{a}\right)}
\delta(\Omega-\Xi)
\delta^2({\bf k_\perp}-{\bf l_\perp}),
\\ \label{bb9}
{}_\text{M}\bra{0}\bi[\Omega{\bf k_\perp}] \bi[\Xi{\bf l_\perp}]\d \ket{0}_\text{M}
=&
\frac{e^{\frac{\pi\Omega}{a}}}{2\sinh\left(\frac{\pi\Omega}{a}\right)}
\delta(\Omega-\Xi)
\delta^2({\bf k_\perp}-{\bf l_\perp}),
\\
{}_\text{M}\bra{0}\bi[\Omega{\bf k_\perp}] \bii[\Xi{\bf l_\perp}] \ket{0}_\text{M}
=&
\frac{1}{2\sinh\left(\frac{\pi\Omega}{a}\right)}
\delta(\Omega-\Xi)
\delta^2({\bf k_\perp}+{\bf l_\perp}),
\\
{}_\text{M}\bra{0}\bi[\Omega{\bf k_\perp}]\d \bii[\Xi{\bf l_\perp}] \ket{0}_\text{M}
=&
0,
\\
{}_\text{M}\bra{0}\bi[\Omega{\bf k_\perp}] \bii[\Xi{\bf l_\perp}]\d \ket{0}_\text{M}
=&
0.
\end{align}
The first three hold also with the index I replaced with II.
The remaining ones may obtained from these by complex conjugation.

Now we can compute the covariance matrix elements.
Starting with the upper-left $2\times 2$ block:
\begin{align}
\left(\sigma^d_{\text{vac}}\right)_{11}
=&
2\Re {}_\text{M}\bra{0}\hat{d}_\text{I}\hat{d}_\text{I} + \hat{d}_\text{I}\hat{d}_\text{I}\d \ket{0}_\text{M} \nonumber
\\=&
2\Re\iint\text{d}\Omega\text{d}\Xi
\iint\text{d}^2{\bf k_\perp}\text{d}^2{\bf l_\perp} \nonumber
\oI
\\&
\big[
\oIl{}_\text{M}\bra{0}\bi[\Omega{\bf k_\perp}] \bi[\Xi{\bf l_\perp}] \ket{0}_\text{M}+
\oIl\s {}_\text{M}\bra{0}\bi[\Omega{\bf k_\perp}] \bi[\Xi{\bf l_\perp}]\d \ket{0}_\text{M}
\big] \nonumber
\\=&
\Re\int\text{d}\Omega\int\text{d}^2{\bf k_\perp}
\frac{|\oI|^2}{\sinh\left(\frac{\pi\Omega}{a}\right)}
e^\frac{\pi\Omega}{a} \nonumber
\\=&
1+
\Re\int\text{d}\Omega\int\text{d}^2{\bf k_\perp}
\frac{|\oI|^2}{\sinh\left(\frac{\pi\Omega}{a}\right)}
e^{-\frac{\pi\Omega}{a}}.
\end{align}

The same way we obtain the other elements in the block.
The lower-right block is exactly the same, except that indices I are replaced with II.
Moving on to the off-diagonal blocks:
\begin{align}
\left(\sigma^d_{\text{vac}}\right)_{13}
=&
2\Re {}_\text{M}\bra{0}\hat{d}_\text{I}\hat{d}_\text{II} + \hat{d}_\text{I}\hat{d}_\text{II}\d \ket{0}_\text{M} \nonumber
\\=&
2\Re\iint\text{d}\Omega\text{d}\Xi
\iint\text{d}^2{\bf k_\perp}\text{d}^2{\bf l_\perp}
\oI \nonumber
\\& \label{sigma1318}
\big[
\oII{}_\text{M}\bra{0}\bi[\Omega{\bf k_\perp}] \bii[\Xi{\bf l_\perp}] \ket{0}_\text{M}+
\oII\s {}_\text{M}\bra{0}\bi[\Omega{\bf k_\perp}] \bii[\Xi{\bf l_\perp}]\d \ket{0}_\text{M}
\big] \nonumber
\\=&
\Re\int\text{d}\Omega\int\text{d}^2{\bf k_\perp}
\frac{\oI\oIImk}{\sinh\left(\frac{\pi\Omega}{a}\right)}.
\end{align}

Calculating analogously the other elements of the off-diagonal block, leads to the matrix $N$ given in the Eq. \eqref{noise}, with the definitions \eqref{N1D0}-\eqref{N12D0}.

\subsection{When Rindler wedges do not have a common apex ($D\neq 0$)}\label{Dnon0}

We will first focus on the counter-accelerated case.
The diagonal $2\times 2$ blocks relate to the reduced states of the wedges, thus they may not depend on the separation of the wedges.
They will be the same as in the $D=0$ case, and the only modification will occur in the off-diagonal blocks.
We will compute here one element, and the others are computed analogously.

Combining the decomposition of the field in the Minkowski, and the accelerated frame, we can write:
\begin{equation}
\hat{b}_{\text{I}\Omega{\bf k_\perp}}=
\int\text{d}l_z\int\text{d}^2{\bf l_\perp}
\left(
\alpha_{\Omega{\bf k_\perp}l_z{\bf l_\perp}}^{\text{I}\star}
\hat{a}_{l_z{\bf l_\perp}}
-
\beta_{\Omega{\bf k_\perp}l_z{\bf l_\perp}}^{\text{I}\star}
\hat{a}_{l_z{\bf l_\perp}}\d,
\right)
\end{equation}
and similarly for the wedge II.
This equation has the same form, whether $D=0$ or not, but it should be noted that the Bogolyubov coefficients and the ladder operators appearing on the right hand side, are $D$-dependent. Using this we compute the necessary expectation values:
\begin{align}
{}_\text{M}\bra{0}\bi[\Omega{\bf k_\perp}]\bii[\Xi{\bf n_\perp}]\ket{0}_\text{M}
=&
-\iint\text{d}l_z\text{d}^2{\bf l_\perp}\,\,
\alpha_{\Omega{\bf k_\perp}l_z{\bf l_\perp}}^{\text{I}\star}
\beta_{\Xi{\bf n_\perp}l_z{\bf l_\perp}}^{\text{II}\star},
\\
{}_\text{M}\bra{0}\bi[\Omega{\bf k_\perp}]\bii[\Xi{\bf n_\perp}]\d\ket{0}_\text{M}
=&
\iint\text{d}l_z\text{d}^2{\bf l_\perp}\,\,
\alpha_{\Omega{\bf k_\perp}l_z{\bf l_\perp}}^{\text{I}\star}
\alpha_{\Xi{\bf n_\perp}l_z{\bf l_\perp}}^{\text{II}},
\\
{}_\text{M}\bra{0}\bi[\Omega{\bf k_\perp}]\d\bii[\Xi{\bf n_\perp}]\ket{0}_\text{M}
=&
\iint\text{d}l_z\text{d}^2{\bf l_\perp}\,\,
\beta_{\Omega{\bf k_\perp}l_z{\bf l_\perp}}^{\text{I}}
\beta_{\Xi{\bf n_\perp}l_z{\bf l_\perp}}^{\text{II}\s}.
\end{align}

For $D=0$ the Bogolyubov coefficients take the form~\cite{Crispino2008}:
\begin{align}
\alpha_{\Omega{\bf k_\perp}l_z{\bf l_\perp}}^{\text{I}}
&=
\frac{e^\frac{\pi\Omega}{2a}}{\sqrt{4\pi a\sqrt{l_z^2+l_\perp^2+m^2}\sinh{\frac{\pi\Omega}{a}}}}
\left(
\frac{\sqrt{l_z^2+l_\perp^2+m^2}+l_z}{\sqrt{l_z^2+l_\perp^2+m^2}-l_z}
\right)^{-\frac{i\Omega}{2a}}
\delta^2({\bf k_\perp}-{\bf l_\perp}),
\\
\beta_{\Omega{\bf k_\perp}l_z{\bf l_\perp}}^{\text{I}}
&=-
\frac{e^{-\frac{\pi\Omega}{2a}}}{\sqrt{4\pi a\sqrt{l_z^2+l_\perp^2+m^2}\sinh{\frac{\pi\Omega}{a}}}}
\left(
\frac{\sqrt{l_z^2+l_\perp^2+m^2}+l_z}{\sqrt{l_z^2+l_\perp^2+m^2}-l_z}
\right)^{-\frac{i\Omega}{2a}}
\delta^2({\bf k_\perp}+{\bf l_\perp}),
\\
\alpha_{\Omega{\bf k_\perp}l_z{\bf l_\perp}}^{\text{II}}
&=
\frac{e^\frac{\pi\Omega}{2a}}{\sqrt{4\pi a\sqrt{l_z^2+l_\perp^2+m^2}\sinh{\frac{\pi\Omega}{a}}}}
\left(
\frac{\sqrt{l_z^2+l_\perp^2+m^2}+l_z}{\sqrt{l_z^2+l_\perp^2+m^2}-l_z}
\right)^{\frac{i\Omega}{2a}}
\delta^2({\bf k_\perp}-{\bf l_\perp}),
\\
\beta_{\Omega{\bf k_\perp}l_z{\bf l_\perp}}^{\text{II}}
&=-
\frac{e^{-\frac{\pi\Omega}{2a}}}{\sqrt{4\pi a\sqrt{l_z^2+l_\perp^2+m^2}\sinh{\frac{\pi\Omega}{a}}}}
\left(
\frac{\sqrt{l_z^2+l_\perp^2+m^2}+l_z}{\sqrt{l_z^2+l_\perp^2+m^2}-l_z}
\right)^{\frac{i\Omega}{2a}}
\delta^2({\bf k_\perp}+{\bf l_\perp}).
\end{align}

We facilitate the possible separation between the wedges, we modify the Bogolyubov coefficients exactly in the same manner as in our previous work~\cite{Ahmadi2016a}.
Let us look at the wedge I.
If it is shifted to the right by $\frac{D}{2}$, then equivalently, for simplicity, we may consider shifting the Minkowski coordinates to the left by the same distance.
This would result in $u_{\bf k}(x,y,z,t)\rightarrow u_{\bf k}(x,y,z+\frac{D}{2},t)=e^{i\frac{D}{2}k_z}u_{\bf k}(x,y,z,t)$.
Hence, e.g. $(u_{k_z{\bf k_\perp}}, w_{\text{I}\Omega{\bf l_\perp}})\rightarrow e^{-i\frac{D}{2}k_z}(u_{k_z{\bf k_\perp}}, w_{\text{I}\Omega{\bf l_\perp}})$.
We perform this for all of the Bogolyubov coefficients and find:
\begin{align}\label{modifiedbogos}
\alpha_{\Omega{\bf k_\perp}l_z{\bf l_\perp}}^{\text{I}}& \to e^{-i \frac{D}{2}k_z}\alpha_{\Omega{\bf k_\perp}l_z{\bf l_\perp}}^{\text{I}}
, 
&\beta_{\Omega{\bf k_\perp}l_z{\bf l_\perp}}^{\text{I}} &\to e^{i \frac{D}{2}k}\beta_{\Omega{\bf k_\perp}l_z{\bf l_\perp}}^{\text{I}}
,
\nonumber\\
\alpha_{\Omega{\bf k_\perp}l_z{\bf l_\perp}}^{\text{II}}& \to e^{i \frac{D}{2}k}\alpha_{\Omega{\bf k_\perp}l_z{\bf l_\perp}}^{\text{II}}
, 
&\beta_{\Omega{\bf k_\perp}l_z{\bf l_\perp}}^{\text{II}}& \to e^{-i \frac{D}{2}k}\beta_{\Omega{\bf k_\perp}l_z{\bf l_\perp}}^{\text{II}}.
\end{align}

This altogether leads to the following expression for the covariance matrix element:
\begin{align} \nonumber
\left(\sigma^d_{\text{vac}}\right)_{13}
=&
\frac{1}{2\pi a}\Re
\int\text{d}^2{\bf k_\perp}
\iint\text{d}\Omega\text{d}\Xi
\oI
\\& \nonumber
\Bigg[
\oIIlmk
\frac{e^{\frac{\pi}{2a}(\Omega-\Xi)}}{\sqrt{\sinh\frac{\pi\Omega}{a}\sinh\frac{\pi\Xi}{a}}}
\int\frac{\text{d}l_z}{\sqrt{l_z^2+k_\perp^2+m^2}}
\left(
\frac{\sqrt{l_z^2+k_\perp^2+m^2}+l_z}{\sqrt{l_z^2+k_\perp^2+m^2}-l_z}
\right)^{\frac{i}{2a}(\Omega-\Xi)}
e^{iDl_z}
\\&
+\oIIlk\s
\frac{e^{\frac{\pi}{2a}(\Omega+\Xi)}}{\sqrt{\sinh\frac{\pi\Omega}{a}\sinh\frac{\pi\Xi}{a}}}
\int\frac{\text{d}l_z}{\sqrt{l_z^2+k_\perp^2+m^2}}
\left(
\frac{\sqrt{l_z^2+k_\perp^2+m^2}+l_z}{\sqrt{l_z^2+k_\perp^2+m^2}-l_z}
\right)^{\frac{i}{2a}(\Omega+\Xi)}
e^{iDl_z}
\Bigg].
\end{align}

Again, as in Ref.~\cite{Ahmadi2016a}, we compute the above integrals, to obtain:
\begin{align} \nonumber
\int\frac{\text{d}l_z}{\sqrt{l_z^2+k_\perp^2+m^2}}&
\left(
\frac{\sqrt{l_z^2+k_\perp^2+m^2}+l_z}{\sqrt{l_z^2+k_\perp^2+m^2}-l_z}
\right)^{\frac{i}{2a}(\Omega\pm\Xi)}
e^{iDl_z}
\\&=
2\cosh\frac{\pi(\Omega\pm\Xi)}{2a}
K_{\frac{i(\Omega\pm\Xi)}{a}}(|k_\perp D|)
-
2\frac{\Delta}{|\Delta|}\sinh\frac{\pi(\Omega\pm\Xi)}{2a}
K_{\frac{i(\Omega\pm\Xi)}{a}}\left(\sqrt{k_\perp^2+m^2}\,|D|\right).
\end{align}

Using this result and rearranging the expression, we arrive at the final form of the covariance matrix element:
\begin{align} \nonumber
\left(\sigma^d_{\text{vac}}\right)_{13}
=
\frac{1}{\pi a}&\Re
\int\text{d}^2{\bf k_\perp}
\iint\text{d}\Omega\text{d}\Xi
\frac{\oI}{\sqrt{\sinh\frac{\pi\Omega}{a}\sinh\frac{\pi\Xi}{a}}}
\\ \nonumber
\bigg[&
\oIIlmk
e^{\frac{\pi}{2a}(\Omega-\Xi)(1-\frac{D}{|D|})}
K_{\frac{i(\Omega-\Xi)}{a}}\left(\sqrt{k_\perp^2+m^2}\,|D|\right)
+
\\
&+
\oIIlk\s
e^{\frac{\pi}{2a}(\Omega+\Xi)(1-\frac{D}{|D|})}
K_{\frac{i(\Omega+\Xi)}{a}}\left(\sqrt{k_\perp^2+m^2}\,|D|\right)
\bigg].
\end{align}

From this and the analogous expressions for the other matrix elements we infer the formula \eqref{N12Dnon0}.
In order to proceed with the parallel-accelerated case, we again use the modified Bogolyubov coefficients, but now the ones related to the wedge II are related to $\alpha_{\Omega{\bf k_\perp}l_z{\bf l_\perp}}^{\text{I}}$ differenly because of the reversed orientation of this wedge.
The further steps of the calculation follow along the same lines as shown above.

\subsection{The resulting off-diagonal terms}\label{offdiag}
Here we present resulting off-diagonal terms for $D \neq 0$ in counter-accelerating case:

	\begin{align}   \nonumber
	N_\text{I,II}^\pm
	=
	\frac{1}{\pi a}
	\int\text{d}^2{\bf k_\perp}
	\iint\text{d}\Omega\text{d}\Xi
	\frac{\oI}{\sqrt{\sinh\frac{\pi\Omega}{a}\sinh\frac{\pi\Xi}{a}}}
	\bigg[&
	\oIIlmk
	e^{\frac{\pi}{2a}(\Omega-\Xi)(1-\frac{D}{|D|})}
	K_{\frac{i(\Omega-\Xi)}{a}}\left(\sqrt{k_\perp^2+m^2}\,|D|\right)
	\\ \label{N12Dnon0} 
	\pm&
	\oIIlk\s
	e^{\frac{\pi}{2a}(\Omega+\Xi)(1-\frac{D}{|D|})}
	K_{\frac{i(\Omega+\Xi)}{a}}\left(\sqrt{k_\perp^2+m^2}\,|D|\right)
	\bigg],
	\end{align}
and for parallel-accelerating case:
\begin{align}   \nonumber
N_\text{I,II}^{((\pm}
=
\frac{1}{\pi a}
\int\text{d}^2{\bf k_\perp}
\iint\text{d}\Omega\text{d}\Xi
\frac{\oI}{\sqrt{\sinh\frac{\pi\Omega}{a}\sinh\frac{\pi\Xi}{a}}}
\bigg[&
\oIIlk
e^{\frac{\pi}{2a}\left[(\Omega-\Xi)-(\Omega+\Xi)\frac{D}{|D|}\right]}
K_{\frac{i(\Omega+\Xi)}{a}}\left(\sqrt{k_\perp^2+m^2}\,|D|\right)
\\ \label{N12Dnon0par} 
\pm&
\oIIlk\s
e^{\frac{\pi}{2a}\left[(\Omega+\Xi)-(\Omega-\Xi)\frac{D}{|D|}\right]}
K_{\frac{i(\Omega-\Xi)}{a}}\left(\sqrt{k_\perp^2+m^2}\,|D|\right)
\bigg].
\end{align}

\section{Calculation of the cross elements in the noise matrix for the skew observers\label{computingnoise2}}
In this Appendix we present in detail the calculation and numerical evaluation of the cross elements in the noise matrix for the non-collinear scenario.
For later computational relevance, let us recall time derivatives in both Rindler frames~(\ref{rindler}), expressed in Minkowski coordinates
\begin{align}\label{ders}
\partial_{\eta^{'}}&=a \left\lbrace \left[ (z-D_z) \cos \theta + (y-D_y) \sin \theta \right] \partial_t + t \left[ \cos \theta \partial_z + \sin \theta \partial_y \right]  \right\rbrace   ,\nonumber \\
\partial_{\eta}&= a \left( z \partial_t + t \partial_z \right).
\end{align}

In order to compute the noise matrix elements~(\ref{ns12}), we have to evaluate the scalar products that are in included in Eq.~(\ref{ns12}). For simplicity we introuduce a shorthand notation for the normalization constant of the Rindler mode, $\beta_2(\Omega)=\sqrt{\frac{\sinh\left(\frac{\pi\Omega}{a}\right)}{2\pi^3a}}$. We assume we have a given profile of Bob's wavepacket $\psi_{\text{B}}$ and its Rindler time derivative at $\eta=0$. We revoke at $t=0$:
\begin{align} \label{ns21}
\partial_{\eta'} \psi_{\text{B}} = a \left[  (z-D_z) \cos \theta + (y-D_y) \sin \theta   \right] \partial_t \psi_{\text{B}},
\end{align}
and calculate
\begin{align} \label{ns212}
\Phi_{\text{BI}} =& i \int  \text{d}^2 x \ \left( \psi_{\text{B}}^* \partial_t w_{\text{I} \Omega k} - w_{\text{I} \Omega k} \partial_t \psi_{\text{B}}^*  \right) =  \nonumber \\
&i \int^\infty_{-\infty} \text{d}y \int^\infty_{0} \text{d}z \left(- h_0 \left( \Omega, y,z \right) - h_1 \left(y,z \right)  \right) \beta_2(\Omega) K_{\frac{i\Omega}{a}}\left(\sqrt{k^2+m^2}\,\,z\right)
e^{iky},
\end{align}
where 
\begin{align} \label{ho}
&h_0 \left( \Omega, y,z \right) = \frac{i \Omega}{a z} \psi_{\text{B}}^* (y',z')\nonumber \\
&h_1 \left( y,z \right) = \frac{1}{a \left(  (z-D_z) \cos \theta + (y-D_y) \sin \theta   \right)} \partial_{\eta_B}\psi_{\text{B}}^* (y',z').
\end{align}
Analogously we find that
\begin{align} \label{all1}
\overline{\Phi_{\text{BI}}} &= i \int^\infty_{-\infty} \text{d}y \int^\infty_{0} \text{d}z \left(h_0 \left( \Omega, y,z \right) - h_1 \left(y,z \right)  \right) \beta_2(\Omega) K_{\frac{i\Omega}{a}}\left(\sqrt{k^2+m^2}\,\,z\right) e^{-iky},\nonumber \\
\Phi_{\text{BII}} &= i \int^\infty_{-\infty} \text{d}y \int^0_{-\infty} \text{d}z \left(h_0 \left( \Omega, y,z \right) - h_1 \left(y,z \right)  \right) \beta_2(\Omega) K_{\frac{i\Omega}{a}}\left(-\sqrt{k^2+m^2}\,\,z\right) e^{iky},\nonumber \\
\overline{\Phi_{\text{BII}}} &= i \int^\infty_{-\infty} \text{d}y \int^0_{-\infty} \text{d}z \left(-h_0 \left( \Omega, y,z \right) - h_1 \left(y,z \right)  \right) \beta_2(\Omega) K_{\frac{i\Omega}{a}}\left(-\sqrt{k^2+m^2}\,\,z\right) e^{-iky}.
\end{align}
With this, we can write down the expressions for $N_\text{AB}^{\pm}$ in a more condensed form:
\begin{align} \label{nabs}
N_\text{AB}^{+} &=-i \int\text{d}\Omega\int\text{d}k \Phi_{\text{AI}}(k) \int^\infty_{-\infty} \text{d}y \int^\infty_{-\infty} \text{d}z \beta_2(\Omega) h_1 \left( y,z \right) K_{\frac{i\Omega}{a}}\left(\sqrt{k^2+m^2}\,\,|z|\right)\times\nonumber \\
& e^{-iky} \frac{1}{\sinh{\frac{\pi \Omega}{a}}}   \times \begin{cases} 
e^{\frac{\pi \Omega}{a}} & z>0 \\
0 & z< 0 
\end{cases},\nonumber \\
N_\text{AB}^{-} &=i \int\text{d}\Omega\int\text{d}k \Phi_{\text{AI}}(k) \int^\infty_{-\infty} \text{d}y \int^\infty_{-\infty} \text{d}z \beta_2(\Omega) h_0 \left( \Omega, y,z \right) K_{\frac{i\Omega}{a}}\left(\sqrt{k^2+m^2}\,\,|z|\right) \times\nonumber \\
& e^{-iky} \frac{1}{\sinh{\frac{\pi \Omega}{a}}}   \times \begin{cases} 
e^{\frac{\pi \Omega}{a}} & z>0 \\
0 & z< 0 
\end{cases}.
\end{align}
We keep $\Phi_{\text{AI}}(k)$ in a short form as it will be simplified by our choice of the shape of the wavefunction.
At this point, to evaluate these expressions, we have to already consider explicit form of the mode function.

In general, we follow the choice from the unrotated case in Section~\ref{modechoice}, but with alternations due to the dimensionality and reasons discussed in Section~\ref{r:2d}.
We keep in mind that Bob's wavepacket is rotated, but its form is the same as Alice's, so $\psi_\text{B}=\psi_\text{A}(y',z')=\psi(y',z')$.
Let us now discuss the choice of the spatial profile of the wavepacket.
We choose to start with an analytical form introduced in Section~\ref{modechoice}:
\begin{align}
\label{modepsi2}
&\psi^\text{a}\Big|_{t=0}=
\mathcal{N}_\psi^\text{a}\,\,
e^{-2(\frac{1}{\mathcal{A} L_\parallel}\log(\mathcal{A}\chi))^2-\frac{2}{L_\perp^2}y^2}
\Im\left\{
I_{-\frac{i\Omega_0}{\mathcal{A}}}\left(\frac{m}{\mathcal{A}}\right)
I_{\frac{i\Omega_0}{\mathcal{A}}}\left(m\,\chi\right)
\right\}
\sin[\kappa_\perp y], \nonumber
\\
&\text{with}\quad\quad 
\partial_\tau\psi^\text{a}\Big|_{\tau=0}=
- i\Omega_0\psi^\text{a}\Big|_{\tau=0}.
\end{align}
Here, superscript a denotes the fact that this wavefunction is analytical and $\mathcal{N}_\psi^\text{a}$ is the normalization constant.
In the unrotated cases, there was not much difference between analytical form and the final wavefunction after the numerical positive-frequency cut-off, but this time this difference will be crucial.
With the analytical formula, we can calculate the overlap with the Rindler mode explicitly:
\begin{align}
\left( \psi_{A}^\text{a} , w_{\text{I}\Omega k} \right) = \left( \frac{\Omega_0}{\A} + \frac{\Omega}{a}  \right) \mathcal{N}_\psi^\text{a} \beta_2(\Omega) i L_\perp \frac{\pi}{2} e^{-\frac{1}{8} L_\perp^2 (k^2+\kappa_\perp^2)} \sinh{\left( \frac{1}{4} L_\perp^2 k \kappa_\perp \right) }    g\left( \Omega, |k| \right),
\end{align}
where 
\begin{align} \label{geez}
g\left( \Omega, k \right) =  \int_0^{\infty}\frac{\text{d}\chi}{\chi} \, K_{\frac{i \Omega}{a}} \left( \sqrt{k^2 + m^2}  \chi \right)\Im\left\{
I_{-\frac{i\Omega_0}{\A}}\left(\frac{m}{\A}\right)
I_{\frac{i\Omega_0}{\A}}\left(m\,\chi\right)
\right\}  e^{-2\left( \frac{1}{\A L_\parallel}\log{(\A \chi)}  \right)^2 }.
\end{align}

For numerical reasons we have to choose the wavepacket that is relatively big and close to horizon, namely described by parameters $\A=2.0$, $L_\parallel=0.5$, $L_\perp=0.1$, $\Omega_0=5.0$, $\kappa_\perp=2.0$.
As it was mentioned earlier, we project our mode function onto positive frequencies:
\begin{align} \label{pack}
\psi&=\mathcal{N}_\psi \int\text{d}\Omega\int\text{d}k \left( \psi^\text{a} , w_{\text{I}\Omega k} \right) w_{\text{I}\Omega k},\nonumber \\
\partial_{\eta}\psi&=-i\mathcal{N}_\psi \int\text{d}\Omega\int\text{d}k \  \Omega \left( \psi_{A}^\text{a} , w_{\text{I}\Omega k} \right) w_{\text{I}\Omega k},
\end{align}
where $\mathcal{N}_\psi$ is the new normalization constant.
The second equation comes from the differentiation of the Rindler mode.
It is noteworthy that the structure of positive frequency overlaps doesn't change
\begin{align} \label{sadas}
\Phi_{\text{AI}}=\mathcal{N}_\psi \left( \psi^\text{a} , w_{\text{I}\Omega k} \right).
\end{align}

In general, we could choose any function acting as $\left( \psi^\text{a} , w_{\text{I}\Omega k} \right)$ to create a wavefunction by the above prescription.
However, the localization of this mode function is a very subtle thing -- in Eq.~(\ref{pack}) the Rindler modes have to interfere constructively in a finite region of space and destructively anywhere else.
That's why we have chosen an analytical form that we know to be localized to start our considerations.
However, the spectrum of such a function, even for some drastic parameters, was not suitable for our calculations, as it occupied also large values of $\Omega$.
Therefore, we tried to squeeze this spectrum by rescaling function $ g\left( \Omega, k \right)$ to fit into smaller values of $\Omega$ and $k$ and hope for wavepacket and its derivative to be localized in space.
The result of this action is the wavepacket depicted in Fig.~(\ref{wv}).
From now, by $ g\left( \Omega, k \right)$ we mean the rescaled version.

With this, we arrive at the final expressions for the noise matrix elements:
\begin{align} \label{pl}
N_\text{AB}^{+} &= \left( \frac{\mathcal{N}_\psi}{\mathcal{N}_\psi^\text{a}}\right)^2 \mathcal{N}_\psi^\text{a} i L_\perp \frac{\pi}{2} e^{-\frac{1}{8} L_\perp^2 \kappa_\perp^2} \frac{1}{2 \pi^3 a} \int \text{d}\Omega \left( \frac{\Omega_0}{\A} + \frac{\Omega}{a}  \right) \int \text{d}k e^{-\frac{1}{8} L_\perp^2 k^2} \sinh{\left( \frac{1}{4} L_\perp^2 k \kappa_\perp \right) }\times \nonumber \\
& g\left( \Omega, |k| \right) 
\int \text{d}z \  K_{\frac{i\Omega}{a}}\left(\sqrt{k^2+m^2}\,\,|z|\right)  \times
\nonumber \\
&\begin{cases} 
e^{\frac{\pi \Omega}{a}} & z>0 \\
0 & z< 0 
\end{cases} \times \mathcal{F}_y{\left\lbrace \frac{\partial_{\eta} \psi_{\text{B}}^* (y',z')}{ a \left(  (z-D_z) \cos \theta + (y-D_y) \sin \theta   \right)} \right\rbrace } (k),
\end{align}
\begin{align} \label{min}
N_{AB}^{-} &=- \left( \frac{\mathcal{N}_\psi}{\mathcal{N}_\psi^\text{a}}\right)^2 \mathcal{N}_\psi^\text{a} i L_\perp \frac{\pi}{2} e^{-\frac{1}{8} L_\perp^2 \kappa_\perp^2} \frac{1}{2 \pi^3 a} \int \text{d}\Omega \ \Omega \left( \frac{\Omega_0}{\A} + \frac{\Omega}{a}  \right) \int \text{d}k e^{-\frac{1}{8} L_\perp^2 k^2} \sinh{\left( \frac{1}{4} L_\perp^2 k \kappa_\perp \right) }\times \nonumber \\
&   g\left( \Omega, |k| \right) 
\int \text{d}z \  K_{\frac{i\Omega}{a}}\left(\sqrt{k^2+m^2}\,\,|z|\right)  
\times \nonumber \\
& \begin{cases} 
e^{\frac{\pi \Omega}{a}} & z>0 \\
0 & z< 0 
\end{cases} \times \mathcal{F}_y{\left\lbrace \frac{i \psi_{\text{B}}^* (y',z')}{ a z} \right\rbrace } (k).
\end{align}
where $\mathcal{F}_y$ denotes Fourier transform with respect to the variable $y$.
They both consist of quadruple integrals, but realization that they can be expressed in terms of Fourier transform allows to use a Fast Fourier Transform algorithm that greatly speeds up the calculations.

\end{document}